 \documentclass[final,3p,times]{elsarticle}

\usepackage{amssymb}
\usepackage{subfigure}
\usepackage{amsmath}
\usepackage{pifont}
\usepackage{bbding}
\usepackage{algorithm}
\usepackage{algorithmic}
\usepackage{multirow}
\usepackage{float}

\journal{Artificial Intelligence}

\begin{document}

\begin{frontmatter}



\title{ITEACH-Net: Inverted Teacher-studEnt seArCH Network for Emotion Recognition in Conversation}


\author[label1,label2]{Haiyang Sun\corref{cor1}}
 \ead{sunhaiyang2021@ia.ac.cn}
 
\author[label2]{Zheng Lian\corref{cor1}}
\ead{lianzheng2016@ia.ac.cn}

\author[label2,label3]{Chenglong Wang}

\author[label2,label4]{Kang Chen}

\author[label1,label2]{Licai Sun}

\author[label1,label2]{Bin Liu\corref{cor2}}
\ead{liubin@nlpr.ia.ac.cn}

\author[label5]{Jianhua Tao\corref{cor2}}
\ead{jhtao@tsinghua.edu.cn}

\cortext[cor1]{Equal contribution.}
\cortext[cor2]{Corresponding authors.}

\affiliation[label1]{organization={School of Artificial Intelligence, University of Chinese Academy of Sciences},
            city={Beijing},
            country={China}}
            
\affiliation[label2]{organization={Institute of Automation, Chinese Academy of Sciences},
	city={Beijing},
	country={China}}
	
\affiliation[label3]{organization={University of Science and Technology of China},
	city={Anhui},
	country={China}}
	
\affiliation[label4]{organization={Peking University},
	city={Beijing},
	country={China}}
	
\affiliation[label5]{organization={Department of Automation,Tsinghua University},
	city={Beijing},
	country={China}}

\begin{abstract}
Due to the widespread applications of conversations in human-computer interaction, Emotion Recognition in Conversation (ERC) has attracted increasing attention from researchers. 
Although many studies have achieved notable results, there remain two critical challenges that hinder the development of ERC. Firstly, there is a lack of exploration into mining deeper insights from the data itself for conversational emotion tasks. Secondly, the systems exhibit vulnerability to random modality feature missing, which is a common occurrence in realistic settings.
Focusing on these two key challenges, we propose a novel framework for incomplete multimodal learning in ERC, called "Inverted Teacher-studEnt seArCH Network (ITEACH-Net)." ITEACH-Net comprises two novel components: the \textit{Emotion Context Changing Encoder} (ECCE) and the \textit{Inverted Teacher-Student} (ITS) framework. 
Specifically, leveraging the tendency for emotional states to exhibit local stability within conversational contexts, ECCE captures these patterns and further perceives their evolution over time.
Recognizing the varying challenges of handling incomplete versus complete data, ITS employs a teacher-student framework to decouple the respective computations. Subsequently, through Neural Architecture Search, the student model develops enhanced computational capabilities for handling incomplete data compared to the teacher model.
During testing, we design a novel evaluation method, testing the model's performance under different missing rate conditions without altering the model weights. We conduct experiments on three benchmark ERC datasets, and the results demonstrate that our ITEACH-Net outperforms existing methods in incomplete multimodal ERC. 
We believe ITEACH-Net can inspire relevant research on the intrinsic nature of emotions within conversation scenarios and pave a more robust route for incomplete learning techniques. Codes will be made available.

\end{abstract}

%


\end{frontmatter}


\section{Introduction}

Emotion Recognition in Conversation (ERC) aims to analyze the emotional states of speakers in conversational contexts. It has found extensive applications in areas such as social media analysis and human-computer interaction \cite{park2016multimodal, li2022bieru, lian2022pirnet}. Recent studies have achieved promising results in ERC \cite{chudasama2022m2fnet, hazarika2021conversational, lian2022smin}. However, there are still some limitations that need to be addressed. 

First, unlike sentence-level emotion understanding tasks, conversations feature longer sequences and important context information. Long-sequence conversations often entail complex context changes, the model need to carefully consider these changes to better understand conversation data. While many works have strengthened the ability of models to process context information \cite{shou2023comprehensive, CTNet, song2022multi, yingjian2023emotionic}, these methods are general in conversational tasks. They have not deeply considered the unique patterns of emotional expression in conversations, and there is an inability to leverage such prior knowledge to fully exploit data.

Second, real-world scenarios introduce various challenges that impact the collection of multimodal data. Factors such as background noise in speech, obstacles in images, and recognition errors in text may result in the omission of modal information, introducing uncertainty to the completeness of the data. Consequently, the model must exhibit robustness in addressing incomplete data. 
Although many works have performed well in incomplete learning \cite{MMIN, yuan2021transformer, GCNet, EMT-DLFR}, they overlook the difference in computational power requirements between models for incomplete data tasks and complete data tasks, failing to endow models with sufficient capability to handle missing data.

Third, the disturbances data encounters are uncertain. Therefore, when evaluating the robustness of models, it is essential to consider their performance in scenarios where the incompleteness of the data dynamically changes. Existing evaluation methods, which involve training model weights separately for testing under various missing rate scenarios \cite{SelfMM, MMIF, zuo2023exploiting, TFRNet}, fail to authentically reflect the model's robustness.

\begin{figure*}[t]
	\centering
	\includegraphics[width=0.95\linewidth]{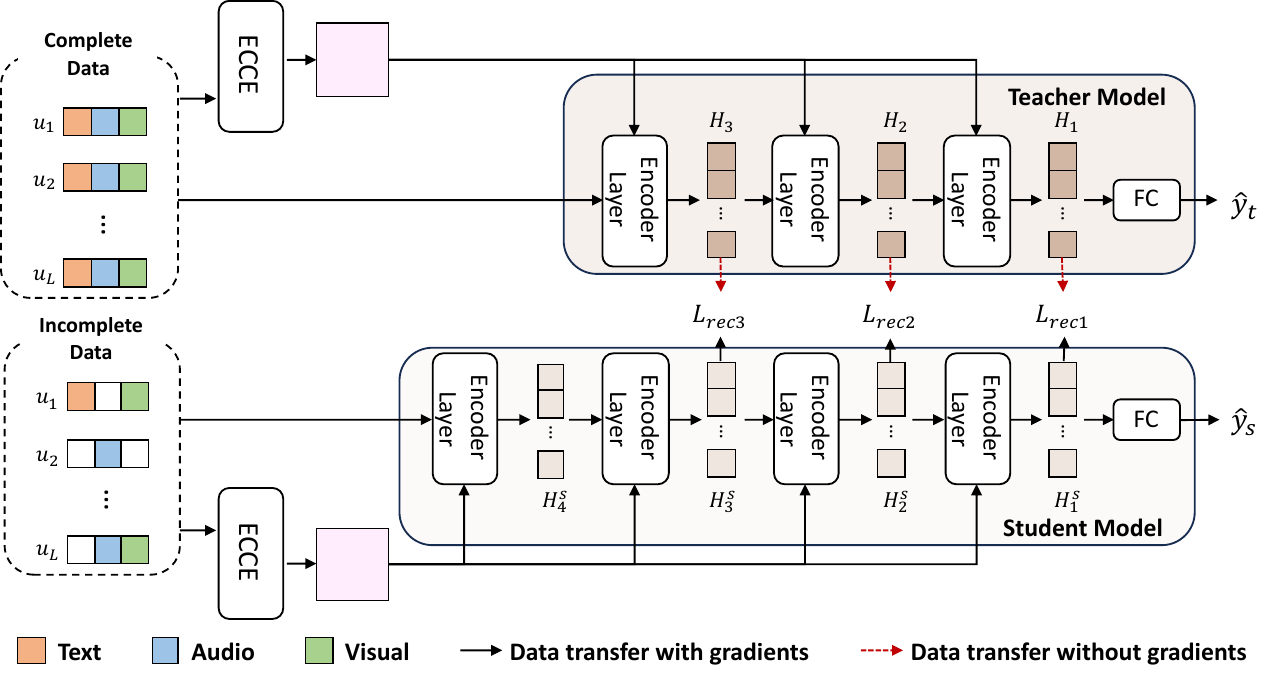}
	
	\caption{The overall structure of Inverted Teacher-studEnt seArCH Network (ITEACH-Net) with the trimodal setting. The \textbf{Inverted Teacher-Student} framework employs a complex student model to learn from a simple teacher model. The \textbf{Emotion Context Changing Encoder} (ECCE) captures the intricate context information within conversations.
	}
	\label{macro_structure}
\end{figure*}

In this paper, we propose a novel framework for incomplete learning in ERC, called "Inverted Teacher-studEnt seArCH Network (ITEACH-Net)". Figure \ref{macro_structure} illustrates the overall structure of our method. To better understand the unique patterns in emotional conversations, as described in Figure \ref{conversation_trend}, we design a novel encoder based on the context changes that enables the model to simultaneously consider them from both local and global perspectives, called Emotion Context Change Encoder (ECCE). To enhance the model's ability to handle missing data, we propose a novel training framework, called Inverted Teacher-Student (ITS), which allows the student model to utilize computationally intensive techniques, such as Neural Architecture Search (NAS) \cite{DARTS, burgerformer, EmotionNAS, SERMFAS}, to acquire knowledge from the teacher trained on complete data, thereby enhancing its capability in processing incomplete data. 

To verify the actual robustness of models, we propose a novel evaluation method to simulate dynamic data missing scenarios and conduct experiments on three benchmark ERC datasets.
Specifically, we fix the model weights and assess its performance across different missing rates.

Through quantitative and qualitative analysis, we demonstrate that our ITEACH-Net outperforms currently advanced approaches. The main contributions of this paper are summarized as follows:

\begin{itemize}
	
	\item We introduce a focus on local stable emotional patterns within conversational data, and design a novel encoding method, called Emotion Context Change Encoder, that enables the model to simultaneously consider them from both local and global perspectives.
	
	\item We propose a novel Inverted Teacher-Student training method as the foundational framework for incomplete data, which decouples the computation of complete and incomplete data, and leverages NAS to endow the student model with stronger encoding capabilities, thereby improving its robustness in the face of incomplete data.
	
	\item Unlike existing works that train individual models for specific missing rates, we propose a novel evaluation approach, namely assessing the average performance of a unified model across all missing rates. This evaluation method is closer to the dynamic data missing scenario and fills the gap in current research.
	
	\item Experimental results on three benchmark datasets demonstrate that our method, ITEACH-Net, is superior to existing state-of-the-art approaches in ERC.
	
\end{itemize}

\begin{figure*}[t]
	\centering
	\includegraphics[width=0.9\linewidth]{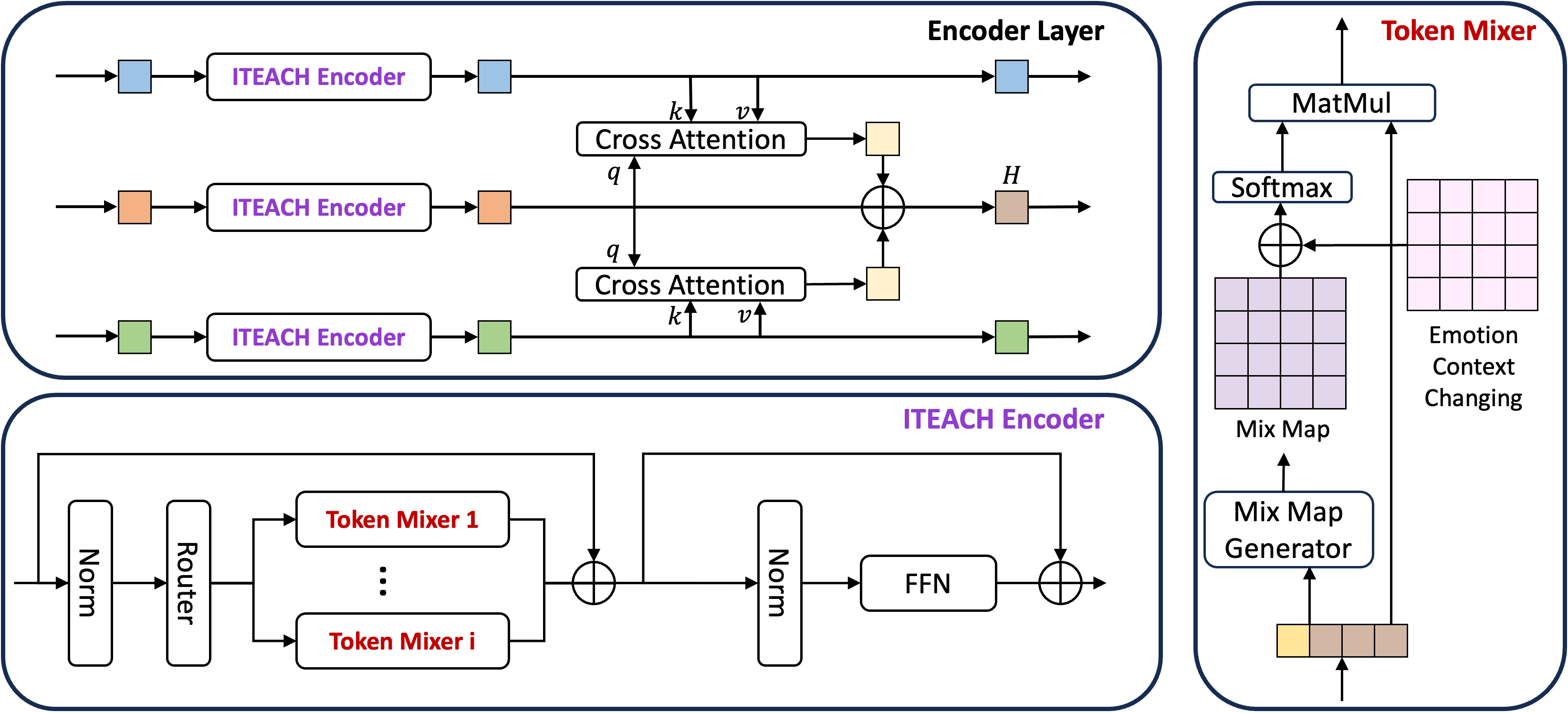}
	
	\caption{The computational modules within the \textbf{Teacher Model} and the \textbf{Student Model}.}
	\label{encoder_layer}
\end{figure*}

\section{Related Work}

\subsection{Context Modeling in Conversation}
In comparison to sentence-level emotion recognition tasks, context information in conversations is extremely crucial \cite{shou2023comprehensive}. Utilizing temporal modeling methods to consider context information is simple and effective, whether using positional encoding \cite{ishiwatari2020relation} or temporal models \cite{lian2019conversational, lian2020conversational, CTNet}, both can do this. Based on these methods, many works have further enhanced the modeling of context information. Yang et al. \cite{yang2022contextual} introduce context information by embedding previous statements between interlocutors, thereby enhancing the emotional representation of the current utterance. Liu et al. \cite{liu2022multi} use a BLSTM \cite{LSTM} layer to learn long-term dependencies and utterance-level context information, and a multi-head self-attention layer to make the model focus on the features most related to the emotions. Song et al. \cite{song2022multi} innovatively introduce the concept of global emotional atmosphere, using the most dominant emotion in the conversation as guiding information, and use graph neural networks \cite{bruna2014spectral}  and GRUs \cite{cho2014properties} in a concatenated framework to consider local and long-term context information. However, these context modeling methods are universally applicable across various conversational tasks, they do not delve into the unique characteristics inherent in conversational emotional data.

Further, Liu et al. \cite{yingjian2023emotionic} consider context modeling from the perspectives of emotional inertia and contagion, designing two independent modules to capture identity-based global context dependencies and extract speaker- and temporal-aware local context information. Tu et al. \cite{tu2022context} enhance the model's ability to capture emotional information by starting from the dependency relationship between commonsense knowledge and context. Peng et al. \cite{peng2022modeling} explore the interrelationship between intention, action, and emotion, while Ai et al. \cite{ai2024gcn} use graph neural networks to model the dependencies between speakers and events in conversations. These methods, although considering the characteristics of emotional information in conversational tasks, primarily focus on the association with external knowledge and lack an in-depth exploration of the intrinsic properties of the data itself.

\subsection{Incomplete Multimodal Learning}

Learning from incomplete multimodal data is a fundamental research area in machine learning. One simple approach involves data imputation, followed by utilizing existing classification methods on the imputed data. Adding zero vectors or using the mean values of the same modality within a class as missing data are the most straightforward methods \cite{Parthasarathy_Sundaram_2020, Zhang_Cui_Han_Zhou_Fu_Hu_2020, CPMNet}. However, the performance of such approaches inevitably reaches an upper limit, as they cannot leverage deeper multimodal correlations to improve downstream task performance. Consequently, some studies start leveraging the correlations between different modalities to optimize this process, manifesting as a low-rank data matrix resulting from high correlation \cite{Yang_Yumer_Asente_Kraley_Kifer_Giles_2017, Liang_Liu_Tsai_Zhao_Salakhutdinov_Morency_2019}. Many works, therefore, project the data into a common space using low-rankness \cite{Fan_Chen_Guo_Zhang_Kuang_2017, Liang_Liu_Tsai_Zhao_Salakhutdinov_Morency_2019}. These methods have shown promising results in incomplete multimodal learning.

However, with the advancement of deep learning, methods based on deep learning have significantly surpassed the aforementioned approaches. For instance, Duan et al. \cite{Yanjie_Duan_Yisheng_Lv_Wenwen_Kang_Yifei_Zhao_2014} utilize autoencoders to estimate missing data, achieving commendable results in incomplete multimodal learning. Tran et al.\cite{CRA} further propose the Cascaded Residual Autoencoder (CRA) to enhance the imputation capability of autoencoders. Additionally, Zhao et al. \cite{MMIN} combine CRA with cycle-consistency loss for cross-modal imputation, outperforming existing methods. Furthermore, Zuo et al. \cite{zuo2023exploiting} and Sun et al. \cite{EMT-DLFR} enhance the imputation ability of autoencoders for missing data from both unimodal and multimodal perspectives, or both high- and low-level features, achieving outstanding performance. Meanwhile, Lian et al. \cite{GCNet} delve deeper into considering temporal and speaker information in conversations, achieving state-of-the-art performance in ERC. 
Although these works have shown excellent performance in handling incomplete data, using the same network structure to model both complete and incomplete data means that the model lacks additional capabilities to strengthen learning from incomplete data.


\section{Method}
\subsection{Input Format}

In this task, a conversation $C = { \{(u_i, y_i)\}}^{L}_{i=1}$ consists of $L$ utterances, and the $i^{th}$ utterance $u_i$ with a label $y_i$. For each utterance, we extract features $x_m \in \mathbb{R}^{1 \times d_m}$, $m \in \{t, a, v\}$, from text, audio and visual modalities, where $d$ represents the feature dimension, and concatenate them in the order of speaking to obtain $X_m \in \mathbb{R}^{L \times d_m}$, to represent the entire conversation. 

\subsection{Inverted Teacher-Student}
To ensure that the model achieves performance akin to that with complete data when processing incomplete information, existing approaches primarily focus on encoding the incomplete data into deep representations to approximate those derived from complete data. This introduces additional computational burden to the model in a multitask format, necessitating greater computational capacity than models operating on complete data.
Therefore, we introduce the Teacher-Student framework to decouple these two computations, where the teacher model is trained on complete data, and the student model focuses on utilizing incomplete data to emulate the teacher model's performance. Furthermore, we design an Inverted Teacher-Student framework, providing the student model with a more complex structure than the teacher, enabling it to possess sufficient computational capacity to follow the teacher's performance.

Teacher-Student training framework is designed to distill knowledge from a complex teacher model into a simpler student model \cite{hinton2015distilling, touvron2021training, li2024kd}, Inverted Teacher-Student serves as a framework allowing a complex student to emulate the performance of a simple teacher. Our framework is illustrated in Figure \ref{macro_structure}.

We define complete multimodal features as $\{X_t, X_a, X_v\}$ and incomplete multimodal data as $\{X_t^s, X_a^s, X_v^s\}$. We first capture their emotion context changing information: 
\begin{equation}
	\{E_t, E_a, E_v\} = \mathrm{ECCE}(\{X_t, X_a, X_v\})
\end{equation}
\begin{equation}
	\{E_t^s, E_a^s, E_v^s\} = \mathrm{ECCE_s}(\{X_t^s, X_a^s, X_v^s\})
\end{equation}
where $E_m \in \mathbb{R}^{L \times L}$. Subsequently, they pass through multiple Encoder Layers in their respective models, generating hidden features under the guidance of $E_m$:
\begin{equation}
	\{H_i\}_{i=1}^{L_{tea}} = \mathrm{Teacher}(\{X_t, X_a, X_v\}, \{E_t, E_a, E_v\})
\end{equation}
\begin{equation}
	\{H_j^s\}_{j=1}^{L_{stu}} = \mathrm{Student}(\{X_t^s, X_a^s, X_v^s\}, \{E_t^s, E_a^s, E_v^s\})
\end{equation}
$L_{tea}$ and $L_{stu}$ denote the numbers of Encoder Layers for the Teacher Model and Student Model, respectively. \textbf{It's worth noting that, for ease of understanding, we label the hidden feature closest to the classification layer as $H_1$.} To enhance the computing capability of the Student Model, we allocate more Encoder Layers to it.

In the end, the final layer of hidden features $H_1$ and $H_1^s$ is employed by the classifier to predict emotion labels. Simultaneously, the masked portions of the last three layers of hidden features from the Student Model, $\{H_1^s, H_2^s, H_3^s\}$, are computed for distance loss, aiming to acquire the knowledge from the Teacher Model, $\{H_1, H_2, H_3\}$.

\subsection{Emotion Context Changing Encoder}

\begin{figure}[t]
	\centering
	\includegraphics[width=0.6\linewidth]{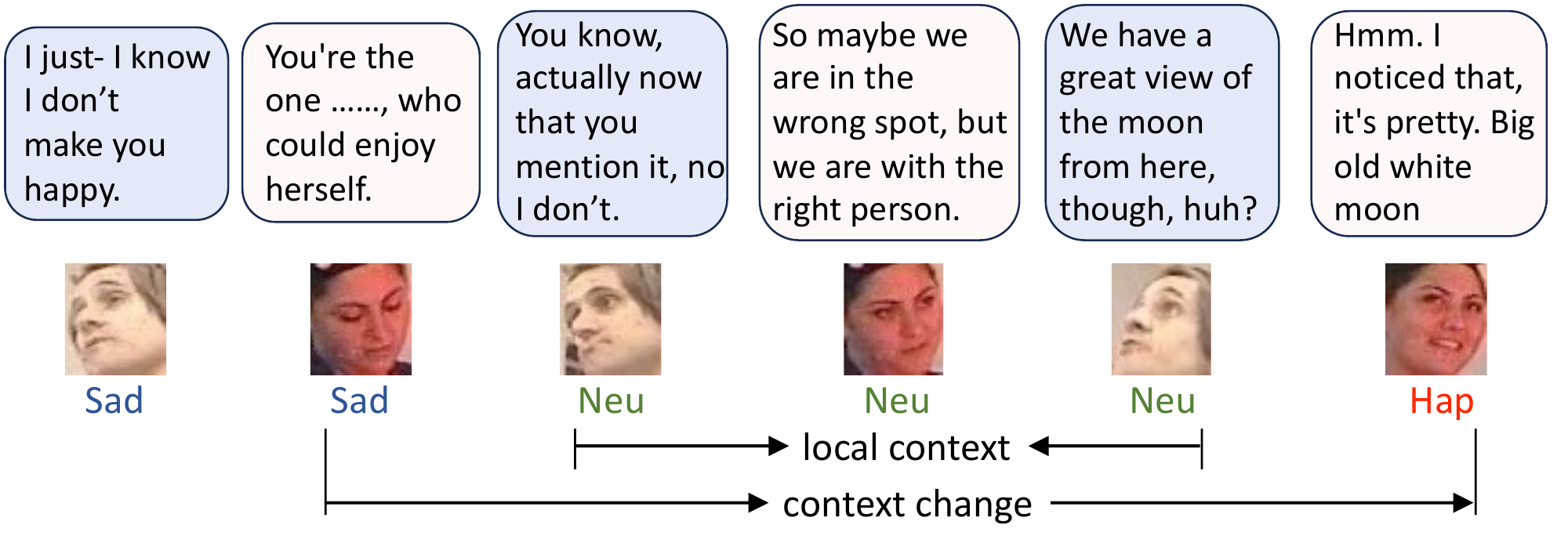}
	
	\caption{In conversations, the speakers' states tend to maintain a relatively stable pattern within the local context. As the context change, this pattern evolves.}
	\label{conversation_trend}
\end{figure}

As depicted in Figure \ref{conversation_trend}, the emotional state of speakers in a conversation tends to maintain a stable pattern locally, which evolves with context changes. To enhance model's perception of these dynamics, we design a novel encoding method, Emotion Context Changing Encoder (ECCE). The encoder consists of two specific steps: encoding local emotion context and encoding global context changing.

\subsubsection{Encoding local emotion context}
Given a context window size $w$, for the $i^{th}$ utterance, the calculation process for its local context feature $z^i_m$ can be described as follows:
\begin{equation}
	z^i_m = F_{\text{local}}^m(X_m\left[ i - \lceil \frac{w}{2} \rceil : i + \lfloor \frac{w}{2} \rfloor \right])
\end{equation}
$F_{\text{local}}^m$ represents a replaceable function used for processing features within the window. After calculating for each utterance, we obtain the local context feature $Z_m \in \mathbb{R}^{L \times d_z}$.

\subsubsection{Encoding global context changing}
The local context feature considers only the emotion context within a limited window. However, to account for the dynamics of the context changing over time, further calculations are needed for the $Z_m$. For the context changing feature $e^{i,j}_m$ from the $i^{th}$ to $j^{th}$ utterances, the calculation process can be described as follows:
\begin{equation}
	e^{i,j}_m = F_{\text{global}}^m(Z_m\left[ i : j \right])
\end{equation}
$F_{\text{global}}^m$ represents a replaceable function used for processing context features. During computation, the maximum allowable distance $dist_m$ between $i$ and $j$ can be set to mitigate excessive computational complexity. After calculating for each utterance pair, we obtain the context changing feature $E_m \in \mathbb{R}^{L \times L \times d_e}$.

\begin{algorithm}[t]
	\floatname{algorithm}{Algorithm}
	\setcounter{algorithm}{0}
	\caption{Pseudo code of ECCE.} 
	\hspace*{0.02in} {\bf Input:} $X_m \in \mathbb{R}^{L \times d_m}$, $w$, $dist_m$, $F_{\text{local}}^m$, $F_{\text{global}}^m$ \\
	\hspace*{0.02in} {\bf Output:} $E_m \in \mathbb{R}^{N \times N \times d_{e}}$ \\
	
	\hspace*{0.02in} $FC$ = Linear($d_z, d_e$) \\
	\hspace*{0.02in} $Z_m$ = $F_{\text{local}}^m$($X_m.T$)$.T$ \\
	\hspace*{0.02in} $E_m$ = Zeros($N, N, d_{z}$) \\
	
	\begin{algorithmic}[]
		\FOR{$i$ in $range(0, N)$}
		\FOR{$j$ in $range(i,$ min$(N, dist_m+i))$}
		\STATE $E_m[i][j]$ =  $F_{\text{global}}^m$($Z_m[i:j]$)
		\STATE $E_m[j][i]$ =  $E_m[i][j]$
		\ENDFOR
		\ENDFOR
	\end{algorithmic} 
	\hspace*{0.02in} $E_m$ = $FC$($E_m$) \\
	\hspace*{0.02in} {\bf Return:} $E_m$
	\label{trendcode}
\end{algorithm}

Algorithm \ref{trendcode} presents the pseudocode for the ECCE. In our experiments, we use a 1D convolution operation as $F_{\text{local}}$ and a straightforward average pooling operation as $F_{\text{global}}$.

\subsection{Teacher-Student Model}
\label{encoderlayer}
To enhance the model's perception of complex contexts in conversations, we leverage emotion context changing information $E_m$ to guide the modeling of single-modal sequence information. Additionally, to improve the model's robustness for incomplete data, we propose the use of NAS to enhance the mimetic capabilities of the student model. In this section, we will provide a detailed explanation of these processes. As illustrated in Figure \ref{encoder_layer}, the Encoder Layer and ITEACH Encoder subgraphs depict the model we designed, while the Token Mixer subgraph illustrates the guidance process of $E_m$.

\subsubsection{Encoder Layer}
As shown in Figure \ref{encoder_layer}, multimodal features undergo encoding in their respective ITEACH Encoder at each Encoder Layer. Subsequently, the Cross Attention module merges various modal information into text features. The fused text features will serve as the latent vector $H$ in Figure \ref{macro_structure}. Subsequently, the three modal data will continue to be fed into the next Encoder Layer. The calculation process of Cross Attention can be described as follows:
\begin{equation}
	output = \text{softmax}(\frac{qk^T}{\sqrt{C}})v
\end{equation}
$C$ represents the dimension of $q$. In both rounds of Cross Attention modules, text features consistently serve as $q$, while audio and visual features respectively function as $k$ and $v$ for the two iterations.

\subsubsection{ITEACH Encoder}
\begin{figure}[t]
	\centering
	\includegraphics[width=0.6\linewidth]{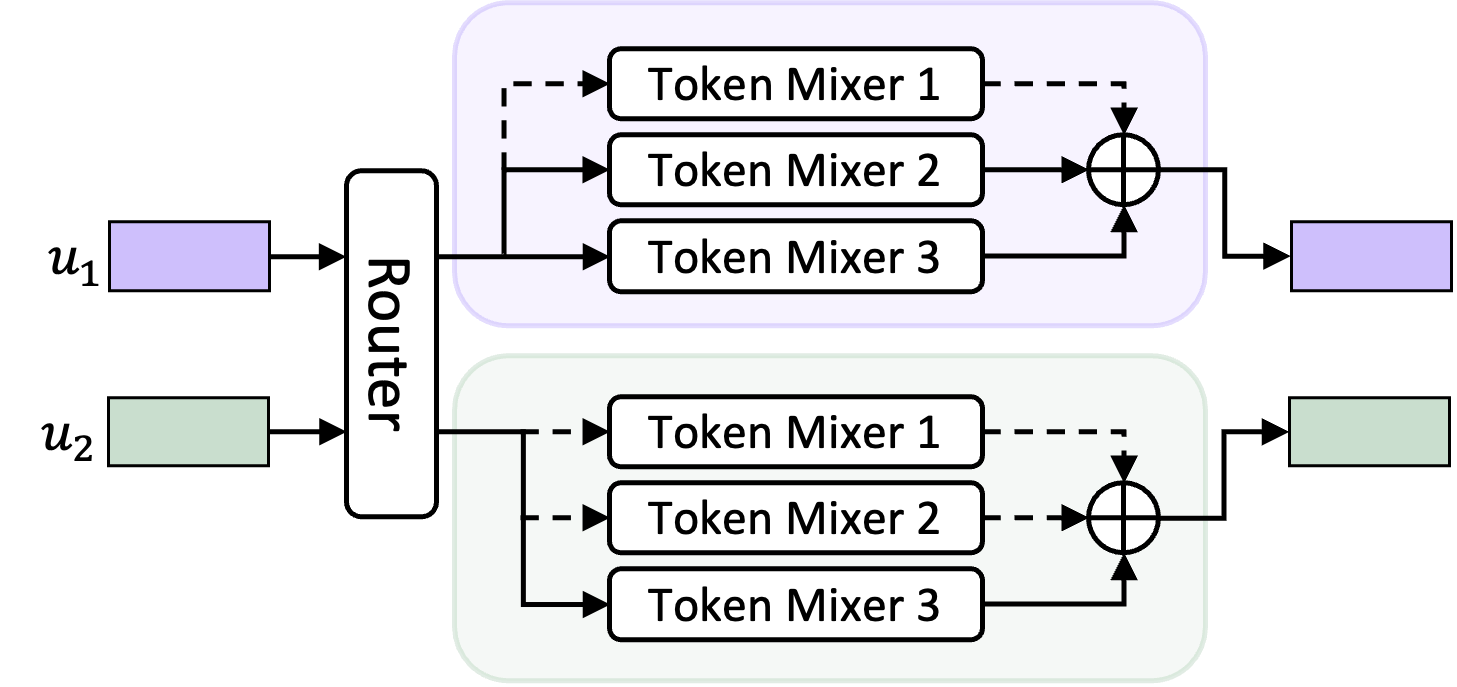}
	
	\caption{When there are three Token Mixers, the Router's search process involves different utterance features selecting different operations. The parameters for Token Mixer i are shared.}
	\label{router_search}
\end{figure}

In the design of the ITEACH Encoder, we retain the overall framework of Metaformer  \cite{metaformer} and conduct a search for the Token Mixer \cite{burgerformer}. Unlike previous workes, to maximize the model's encoding capacity, we utilize the Router module \cite{yu2022efficient} to extend the architecture search process to input-dependent. As shown in Figure \ref{router_search}, different inputs can choose different architectures as needed, rather than opting for a single architecture for all inputs. Given the candidate module set $\mathbb{O}$, where $\mathbb{|O|}$ represents the number of modules, the search process through the Router can be described as follows:
\begin{equation}
	X^m_{\alpha} = \text{softmax}(\text{Router}(X_m))
\end{equation}
\begin{equation}
	output = \sum_{o \in \mathbb{O}}{\sum_{i=1}^{L}{X^m_{\alpha}\left[:, i\right] \times o(X_m\left[i\right])}}
\end{equation}
where $X^m_{\alpha} \in \mathbb{R}^{L\times |\mathbb{O}|}$ and $output \in \mathbb{R}^{L\times d_m}$. 

While there have been numerous excellent designs for Token Mixers recently, our focus is on the introduction of NAS methods rather than algorithm design. Therefore, we select 4 simple and lightweight operations as candidates. Please refer to the appendix for a detailed description of them.

In our experiments, when the Router is a linear layer and the Token Mixer comprises the selected four candidates, we denote the ITEACH Encoder as \textbf{NAS}. Conversely, when the Router is an Identity operation and the Token Mixer includes only Self-Attention, we denote it as \textbf{Transformer}.

\subsubsection{Guidance of ECCE}
The role of the Token Mixer is to interact with information from different utterances in conversation. A category of mixers, represented by attention operation, typically generates a Mix Map, which is normalized and used as weights for fusing utterance feature. The $E_m$ serves to adjust the Mix Map, thereby guiding the interaction of sequential information \cite{graphformer}. The specific calculation process can be described as follows:
\begin{equation}
	MixMap_e = MixMap + E_m
\end{equation}
\begin{equation}
	output = X_m \otimes \mathrm{softmax}(MixMap_e)
\end{equation}

\subsection{Joint Optimization}
To simultaneously train the model for both the target task and handling incomplete data, we jointly optimize the emotion loss, $L_{emo}$, and the distance loss, $L_{rec}$. 

Consistent with prior work, we employ cross-entropy loss as the emotion loss for labeled category data. The calculation formula is shown as follows:
\begin{equation}
	L_{emo} = -\frac{1}{2L}\sum_{i=1}^{L}{(y_ilog(\hat{y}_i^t)+y_ilog(\hat{y}_i^s))}
\end{equation}
$y_i \in \mathbb{R}^y$ is the true label. For labeled dimensional value data, we use the mean square error as the emotion loss. The predicted values for this data are transformed into category labels based on a domain range conversion before evaluating classification performance. The loss calculation formula is shown as follows:
\begin{equation}
	L_{emo} = \frac{1}{2L}(||y_i-\hat{y}_i^t||^2+||y_i-\hat{y}_i^s||^2)
\end{equation}
To enable the student model to better emulate the teacher model, we simultaneously calculate the vector distance loss for its last three layers of hidden features. We assign a progressively weighted adjustment to the model's attention to each layer's reconstruction loss:
\begin{equation} 
	L_{rec} = 0.5||H_1^s-H_1||^2+ 0.1||H_2^s-H_2||^2+0.05||H_3^s-H_3||^2
\end{equation}
The hidden layer vectors of the teacher model will be treated as constants during the calculation; therefore, the reconstruction loss will not backpropagate any gradients to the teacher model.

Finally, we merge these two loss functions into a unified objective function. This combined loss is employed to optimize all trainable parameters in an end-to-end manner.
\begin{equation}
	Loss = L_{emo} + L_{rec}
\end{equation}
\section{Experiments}

\subsection{Datasets}

In this section, we provide a comprehensive overview of the dataset's content and the data preprocessing methods employed in our study.

\textbf{CMU-MOSI} \cite{cmumosi} contains 2,199 utterance-level video clips from 93 conversations. Each video clip is annotated with an emotion intensity score ranging from -3 (strongly negative) to 3 (strongly positive).

\textbf{CMU-MOSEI} \cite{cmumosei} is an iterative version of CMU-MOSI and includes 23,453 utterance-level video clips from 3225 conversations with the same annotation format as CMU-MOSI.

\textbf{IEMOCAP} \cite{iemocap} has two annotation formats: a four-class format and a six-class format. The four-class format consists of 5,531 utterance-level video clips from 151 conversations, with each clip labeled as neutral, happy, sad, or angry. The six-class format includes 7,433 utterance-level video clips from 151 conversations, with additional labels for excited and frustrated emotions.

Following \cite{GCNet} \footnote{https://github.com/zeroQiaoba/GCNet}, we extract audio features using wav2vec \cite{wav2vec}, text features using DeBERTa \cite{roberta}, and visual features using MA-Net \cite{manet}. Similar to \cite{GCNet}, taking the case of 70\% data missing as an example, when reading each mini-batch, each modality randomly resets 70\% of the data to zero, but at least one modality is retained for each utterance.

\subsection{Evaluation Metrics}
Following \cite{GCNet}, for CMU-MOSI and CMU-MOSEI, we use MAE as the loss function and select the model based on the best weighted average F1-score (WAF) on the validation set, reporting its WAF on the test set. For IEMOCAP, we adopt a 5-fold leave-one-session strategy and report the mean of the best WAF on the test sets. WAF is computed by excluding the samples with label 0.

\subsection{Evaluation Method}

When evaluating the robustness of a model, it is crucial to consider its performance across various missing scenarios. Previous studies would define different scenarios, assuming that incidental factors hindering data collection would impact data at a fixed rate \cite{MMIF, GCNet, EMT-DLFR}. Consequently, these works masked training and testing data at the same rate to simulate incomplete data scenarios, allowing them train the model and test its performance under this missing rate. This process will repeat with different missing rates to ensure experimental coverage of diverse missing scenarios. In this paper, we refer to this evaluation method as \textbf{Independent Model Evaluation} (IME).

However, in real-world scenarios, the aforementioned assumption is invalid, as the occurrence of incidental factors is uncertain. Therefore, the testing data that the model needs to handle will not maintain a fixed missing rate. To fill this gap, we propose a more realistic evaluation method called \textbf{Unified Model Evaluation} (UME). Once we obtain a well-trained model, we assume it will face a dynamic missing data scenario, where the missing rate of the testing data is no longer fixed.

Specifically, we mask the testing data at a fixed missing rate and then test the model's performance under this missing rate without optimizing its weights. Subsequently, this process will repeat with different missing rates, and the average is taken as a comprehensive performance metric for the robustness.

\subsection{Setup}

We conduct experiments on all datasets using a unified setting, incorporating three modalities: text, audio, and video. The kernelsize of the $F_{local}$ is set to 7, and $dist_m$ is set to 30. Each Encoder Layer has a hidden dimension of 128, an FFN hidden layer ratio of 4.0, and Self-Attention with 8 head nums. The weights of Teacher model and Student model are optimized using AdaW with a learning rate of 5e-4, and Router are optimized using AdaW with a learning rate of 5e-3. The model is trained using a batch size of 32 for 100 epochs. We run each model three times and report average results.
\textbf{It's worth noting that, in this paper, unless otherwise specified, the Teacher model is uniformly set to three Encoder Layers, while the Student model is uniformly set to four.}

\subsection{Baselines}

To evaluate the performance of our proposed ITEACH-Net, we consider the following state-of-the-art incomplete multimodal learning methods as baselines.

\textbf{CPM-Net} \cite{CPMNet} projects all samples into a common space without considering missing patterns. It learns well-structured features by equipping with a clustering-like classification loss.

\textbf{AE} \cite{bengio2006greedy} employs an autoencoder to estimate missing data from partially observed inputs and jointly optimizes the autoencoder's reconstruction loss with the classification loss of the downstream task.

\textbf{CRA} \cite{CRA} combines a series of residual autoencoders into a cascading architecture for data imputation and jointly optimizes imputation and downstream tasks end-to-end.

\textbf{MMIN} \cite{MMIN} combines CRA with cycle consistency learning to predict the latent representations of missing modalities, exhibiting excellent performance under various missing conditions.

\textbf{GCNet} \cite{GCNet} employs graph neural networks to model speaker and context information separately, and uses LSTM to model temporal information, finally jointly optimizing for imputation and downstream tasks. This method outperforms others in incomplete learning for ERC.

\section{Results}

\subsection{Emotion Context Changing Encoder Performance}

\begin{table}[t]
	\centering
	\caption{Ablation results for Emotion Context Change Encoder. The best performance is highlighted in bold.}
	\begin{tabular}{c|cc|c}
		\hline \hline
		Dataset                                                                         & Local      & Global     & WAF$\uparrow$   \\ \hline \hline
		\multirow{3}{*}{\begin{tabular}[c]{@{}c@{}}IEMOCAP\\ (four-class)\end{tabular}} & \ding{53} & \ding{53} & 77.83 \\
		& \ding{53} & \checkmark  & 77.69 \\
		& \checkmark  & \checkmark  & \textbf{78.15} \\ \hline 
		\multirow{3}{*}{\begin{tabular}[c]{@{}c@{}}IEMOCAP\\ (six-class)\end{tabular}}  & \ding{53} & \ding{53} & 60.28 \\
		& \ding{53} & \checkmark  & 59.89 \\
		& \checkmark  & \checkmark  & \textbf{60.66} \\ \hline 
		\multirow{3}{*}{CMU-MOSI}                                                       & \ding{53} & \ding{53} & 85.46 \\
		& \ding{53} & \checkmark  & 85.16 \\
		& \checkmark  & \checkmark  & \textbf{85.74} \\ \hline 
		\multirow{3}{*}{CMU-MOSEI}                                                      & \ding{53} & \ding{53} & 87.31 \\
		& \ding{53} & \checkmark  & 87.12 \\
		& \checkmark  & \checkmark  & \textbf{87.59} \\ \hline \hline
	\end{tabular}
	\label{trendresults}
\end{table}

Recognizing the unique pattern of context changes in Emotion Recognition in Conversational (ERC), we propose a novel encoder called Emotion Context Change Encoder (ECCE). This method takes into account context change information from both local and global perspectives.
To comprehensively analyze the performance of ECCE, we compare the performance of teacher models under different settings, as ECCE is not influenced by data missing. The results are presented in Table \ref{trendresults}, where "Local" denotes the encoding of local emotion context, and "Global" signifies the encoding of global context changes. As the local emotion context information cannot be directly utilized, we are unable to conduct independent performance testing on it.

Firstly, the incorporation of ECCE, derived from two-stage encoding involving Local and Global encoding, exhibits a consistent improvement in model performance across the four tasks compared to vanilla Self-Attention. Simultaneously, forgoing Local stage encoding and directly utilizing utterance features for encoding global context changes results in a decline in model performance across the four tasks.

\begin{figure}[t]
	\centering
	\subfigure[]{
		\includegraphics[width=0.205\linewidth]{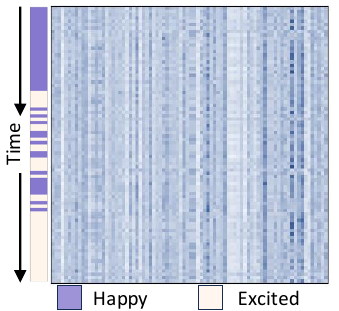}
		\label{attnm}
	}
	\hspace{-0.45cm}
	\subfigure[]{
		\includegraphics[width=0.176\linewidth]{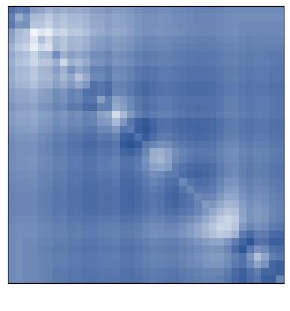}
		\label{onlyglobal}
	}
	\hspace{-0.45cm}
	\subfigure[]{
		\includegraphics[width=0.176\linewidth]{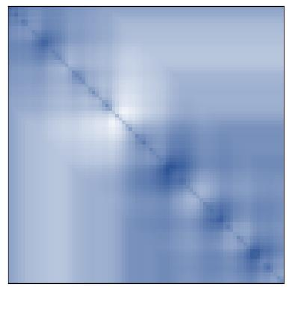}
		\label{trendm}
	}
	
	\caption{Heatmaps generated under different model settings. (a) illustrates the attention map computed by vanilla Self-Attention. (b) illustrates ECCE without encoding local emotion context. (c) illustrates ECCE generated through complete two-stage encoding.}
	\label{TrendMap}
\end{figure}

Furthermore, as illustrated in Figure \ref{TrendMap}, we employ heatmaps for a more intuitive analysis of ECCE. Figure \ref{attnm} depicts the attention map generated by vanilla Self-Attention, Figure \ref{onlyglobal} represents ECCE without Local stage encoding, and Figure \ref{trendm} illustrates the complete ECCE. Darker colors indicate higher weights. All maps originate from the same conversation data, where both speakers initially expressed happiness. As the conversation progressed, a alternation between happy and excited emotions emerged.

As shown in Figure \ref{attnm}, the vanilla Self-Attention, when computing attention maps, tends to allocate higher weights to certain utterances that are more crucial throughout the entire conversation. However, it fails to consider the context in which the two utterances are situated and the changing between them. In contrast, the content in Figure \ref{onlyglobal} demonstrates a certain degree of local attention capability. However, this attention behavior appears somewhat ambiguous and does not accurately discern the emotional context. 

Figure \ref{trendm} illustrates the complete ECCE, which takes into account these aspects effectively. It can be observed that there is a strong local correlation in the top-left and bottom-right corners, indicating that the model can perceive significant differences in the changing context between the two parts of the conversation data. Additionally, there are three more pronounced local correlation regions in the bottom-right, corresponding to three distinct local contextual patterns: frequent emotional changes, a local context of happiness, and a local context of excitement. These results effectively demonstrate the advantage of ECCE in considering changes of emotion contexts in conversations.

\begin{table*}[!h]\small
	\centering
	\caption{Ablation results for Inverted Teacher-Student framework. We report WAF scores(\%), and higher WAF indicates better performance. The Decline metric is obtained by calculating the difference in performance between 0\% and 70\% missing rates. The best performance is highlighted in bold.}
	\label{ITSResults}
	\begin{tabular}{c|c|c|c|ccccccccc}
		\hline \hline
		\multirow{2}{*}{Dataset}                                                        & \multirow{2}{*}{\begin{tabular}[c]{@{}c@{}}\scalebox{0.7}{Training}\\ \scalebox{0.7}{Framework}\end{tabular}} & \multirow{2}{*}{\scalebox{0.8}{Teacher}} & \multirow{2}{*}{\scalebox{0.8}{Student}} & \multicolumn{9}{c}{Missing Rate}                                                                                                                                                                                                                \\ \cline{5-13} 
		&                         &                          &                          & \multicolumn{1}{c|}{0.0$\uparrow$}   & \multicolumn{1}{c|}{0.1$\uparrow$}   & \multicolumn{1}{c|}{0.2$\uparrow$}   & \multicolumn{1}{c|}{0.3$\uparrow$}   & \multicolumn{1}{c|}{0.4$\uparrow$}   & \multicolumn{1}{c|}{0.5$\uparrow$}   & \multicolumn{1}{c|}{0.6$\uparrow$}   & \multicolumn{1}{c|}{0.7$\uparrow$}   & Decline$\downarrow$ \\ \hline \hline
		\multirow{4}{*}{\begin{tabular}[c]{@{}c@{}}IEMOCAP\\ (four-class)\end{tabular}} & PRE                     & TF                       & -                        & \multicolumn{1}{c|}{73.89} & \multicolumn{1}{c|}{72.86} & \multicolumn{1}{c|}{71.52} & \multicolumn{1}{c|}{70.53} & \multicolumn{1}{c|}{69.15} & \multicolumn{1}{c|}{68.65} & \multicolumn{1}{c|}{67.62} & \multicolumn{1}{c|}{65.79} & 8.10   \\
		& ITS                  & TF                       & TF                      & \multicolumn{1}{c|}{77.28} & \multicolumn{1}{c|}{76.26} & \multicolumn{1}{c|}{75.59} & \multicolumn{1}{c|}{74.76} & \multicolumn{1}{c|}{74.00} & \multicolumn{1}{c|}{73.39} & \multicolumn{1}{c|}{72.54} & \multicolumn{1}{c|}{72.06} & 5.22   \\
		& ITS                 & TF                       & NAS                      & \multicolumn{1}{c|}{\textbf{77.47}} & \multicolumn{1}{c|}{\textbf{76.87}} & \multicolumn{1}{c|}{\textbf{76.44}} & \multicolumn{1}{c|}{\textbf{76.02}} & \multicolumn{1}{c|}{\textbf{75.48}} & \multicolumn{1}{c|}{\textbf{74.72}} & \multicolumn{1}{c|}{\textbf{74.19}} & \multicolumn{1}{c|}{\textbf{73.84}} & \textbf{3.63}   \\
		& TS                  & NAS                      & TF                       & \multicolumn{1}{c|}{76.15}      & \multicolumn{1}{c|}{75.48}      & \multicolumn{1}{c|}{74.70}      & \multicolumn{1}{c|}{74.00}      & \multicolumn{1}{c|}{73.37}      & \multicolumn{1}{c|}{72.86} & \multicolumn{1}{c|}{72.33}      & \multicolumn{1}{c|}{71.77}      &   4.38      \\ \hline 
		\multirow{4}{*}{\begin{tabular}[c]{@{}c@{}}IEMOCAP\\ (six-class)\end{tabular}}  & PRE                      & TF                       & -                        & \multicolumn{1}{c|}{56.98} & \multicolumn{1}{c|}{56.02} & \multicolumn{1}{c|}{54.79} & \multicolumn{1}{c|}{53.11} & \multicolumn{1}{c|}{51.67} & \multicolumn{1}{c|}{49.84} & \multicolumn{1}{c|}{47.85} & \multicolumn{1}{c|}{47.46} & 9.52   \\
		& ITS                  & TF                       & TF                      & \multicolumn{1}{c|}{58.59} & \multicolumn{1}{c|}{58.30} & \multicolumn{1}{c|}{57.29} & \multicolumn{1}{c|}{56.79} & \multicolumn{1}{c|}{55.89} & \multicolumn{1}{c|}{55.18} & \multicolumn{1}{c|}{54.74} & \multicolumn{1}{c|}{54.31} & 4.28   \\
		& ITS                 & TF                       & NAS                      & \multicolumn{1}{c|}{\textbf{59.02}} & \multicolumn{1}{c|}{\textbf{58.69}} & \multicolumn{1}{c|}{\textbf{58.53}} & \multicolumn{1}{c|}{\textbf{57.95}} & \multicolumn{1}{c|}{\textbf{57.55}} & \multicolumn{1}{c|}{\textbf{56.75}} & \multicolumn{1}{c|}{\textbf{56.03}} & \multicolumn{1}{c|}{\textbf{55.94}} & \textbf{3.08}   \\
		& TS                  & NAS                      & TF                       & \multicolumn{1}{c|}{58.13}      & \multicolumn{1}{c|}{57.11}      & \multicolumn{1}{c|}{56.41}      & \multicolumn{1}{c|}{55.73}      & \multicolumn{1}{c|}{55.27}      & \multicolumn{1}{c|}{54.50} & \multicolumn{1}{c|}{53.90}      & \multicolumn{1}{c|}{53.16}      &  4.97       \\ \hline 
		\multirow{4}{*}{CMU-MOSI}                                                        & PRE                      & TF                       & -                        & \multicolumn{1}{c|}{84.99} & \multicolumn{1}{c|}{81.65} & \multicolumn{1}{c|}{75.95} & \multicolumn{1}{c|}{75.99} & \multicolumn{1}{c|}{72.27} & \multicolumn{1}{c|}{69.11} & \multicolumn{1}{c|}{62.79} & \multicolumn{1}{c|}{64.66} & 20.33   \\
		& ITS                  & TF                       & TF                      & \multicolumn{1}{c|}{\textbf{85.93}} & \multicolumn{1}{c|}{82.45} & \multicolumn{1}{c|}{79.49} & \multicolumn{1}{c|}{77.21} & \multicolumn{1}{c|}{72.97} & \multicolumn{1}{c|}{69.46} & \multicolumn{1}{c|}{65.68} & \multicolumn{1}{c|}{66.88} & 19.05   \\
		& ITS                 & TF                       & NAS                      & \multicolumn{1}{c|}{84.90} & \multicolumn{1}{c|}{\textbf{83.35}} & \multicolumn{1}{c|}{\textbf{81.04}} & \multicolumn{1}{c|}{\textbf{77.90}} & \multicolumn{1}{c|}{\textbf{76.99}} & \multicolumn{1}{c|}{\textbf{71.94}} & \multicolumn{1}{c|}{\textbf{70.27}} & \multicolumn{1}{c|}{\textbf{68.93}} & \textbf{15.97}   \\
		& TS                  & NAS                      & TF                       & \multicolumn{1}{c|}{85.53}      & \multicolumn{1}{c|}{81.04}      & \multicolumn{1}{c|}{78.31}      & \multicolumn{1}{c|}{75.45}      & \multicolumn{1}{c|}{72.24}      & \multicolumn{1}{c|}{66.26}      & \multicolumn{1}{c|}{62.58}      & \multicolumn{1}{c|}{62.18}      & 23.35        \\ \hline 
		\multirow{4}{*}{CMU-MOSEI}                                                       & PRE                      & TF                       & -                        & \multicolumn{1}{c|}{86.81} & \multicolumn{1}{c|}{85.64} & \multicolumn{1}{c|}{85.01} & \multicolumn{1}{c|}{83.80} & \multicolumn{1}{c|}{82.93} & \multicolumn{1}{c|}{81.89} & \multicolumn{1}{c|}{80.15} & \multicolumn{1}{c|}{79.50} & 7.31   \\
		& ITS                  & TF                       & TF                      & \multicolumn{1}{c|}{\textbf{87.16}} & \multicolumn{1}{c|}{85.86} & \multicolumn{1}{c|}{84.80} & \multicolumn{1}{c|}{83.87} & \multicolumn{1}{c|}{83.12} & \multicolumn{1}{c|}{81.53} & \multicolumn{1}{c|}{80.63} & \multicolumn{1}{c|}{79.42} & 7.74   \\
		& ITS                 & TF                       & NAS                      & \multicolumn{1}{c|}{86.71} & \multicolumn{1}{c|}{\textbf{86.40}} & \multicolumn{1}{c|}{\textbf{84.97}} & \multicolumn{1}{c|}{\textbf{84.40}} & \multicolumn{1}{c|}{\textbf{83.25}} & \multicolumn{1}{c|}{82.00} & \multicolumn{1}{c|}{\textbf{80.65}} & \multicolumn{1}{c|}{\textbf{80.21}} & \textbf{6.50}   \\
		& TS                  & NAS                      & TF                       & \multicolumn{1}{c|}{86.72}      & \multicolumn{1}{c|}{86.22}      & \multicolumn{1}{c|}{84.35}      & \multicolumn{1}{c|}{84.06}      & \multicolumn{1}{c|}{82.52}      & \multicolumn{1}{c|}{\textbf{82.02}}      & \multicolumn{1}{c|}{80.10}      & \multicolumn{1}{c|}{79.04}      &  7.68       \\ \hline \hline
	\end{tabular}
\end{table*}

\subsection{Inverted Teacher-Student Performance}

Dealing with incomplete data poses a significantly greater challenge than handling complete data. However, current methods do not equip models for handling incomplete data with more stronger abilities to follow the performance of models trained with complete data. Therefore, we propose the Inverted Teacher-Student traning framework, which decouples the two computations, and employs a student model with complex structures to handle incomplete data while emulating the performance of a teacher model with simpler structures trained with complete data. To comprehensively analyze the performance of ITS, we compare different training frameworks under different settings. The results are presented in Table \ref{ITSResults}, where the "PRE" represents previous training framework, which uses only one model to handle complete and incomplete data during the training phase. The "TS" training framework represents the vanilla Teacher-Student framework, where the teacher model is more complex than the student model. "TF" is an abbreviation for Transformer. \textbf{It's worth noting that, in the TS training framework, the teacher model is equipped with four encoder layers, while the student model is equipped with three.}

Firstly, the performance of the Transformer under PRE training framework is not satisfactory. It not only fails to achieve outstanding performance with complete data but also exhibits poor capability in handling incomplete data. The IST training framework brings about a noticeable improvement in performance. The student model, guided by a simpler teacher model, not only performs better with complete data, surpassing the baselines across all four tasks, but also demonstrates significant performance improvements in handling incomplete data. In transitioning from complete data to a severe missing scenario with a 70\% data loss, the ITS training framework exhibits less performance degradation. Compared to the PRE training framework, ITS incurs a smaller loss of 2.88\% in the four-class IEMOCAP task, 5.24\% in the six-class IEMOCAP task, and 1.28\% in the CMU-MOSI task. While it shows a slight disadvantage in the CMU-MOSEI task, the difference is negligible due to the similar performance under various missing rates for both frameworks. These results demonstrate the shortcomings of the PRE training framework, while the ITS training framework can better enhance the model's robustness when dealing with incomplete data.

Furthermore, we propose the efficient augmentation of the student model's complexity using Neural Architecture Search (NAS), aiming to enhance its capability in handling incomplete data. While the introduction of NAS doesn't necessarily result in superior performance on complete data, it significantly enhances the model's capability to handle incomplete data. The student model enhanced through NAS achieved the smallest performance decline across all four tasks, with reductions of 3.63\% in the four-class IEMOCAP, 3.08\% in the six-class IEMOCAP, 15.97\% in CMU-MOSI, and 6.50\% in CMU-MOSEI. This substantial surpasses the previous settings.
These results effectively demonstrate that compared to models processing complete data, models handling incomplete data require stronger computational power to process the incomplete data more effectively. Even when trained alongside the same teacher, student models with greater computational capabilities will perform better.

Finally, we present additional results showcasing the performance of a simple student model under the guidance of a complex teacher model. The results indicate that a simple student model struggles to effectively acquire the knowledge from the teacher model, resulting in a more pronounced decline in performance. This comparison highlights the disparity between the ITS and vanilla teacher-student training frameworks in incomplete learning tasks.

\begin{figure*}[t]
	\centering
	\subfigure[IEMOCAP(four-class)]{
		\includegraphics[width=0.4\linewidth]{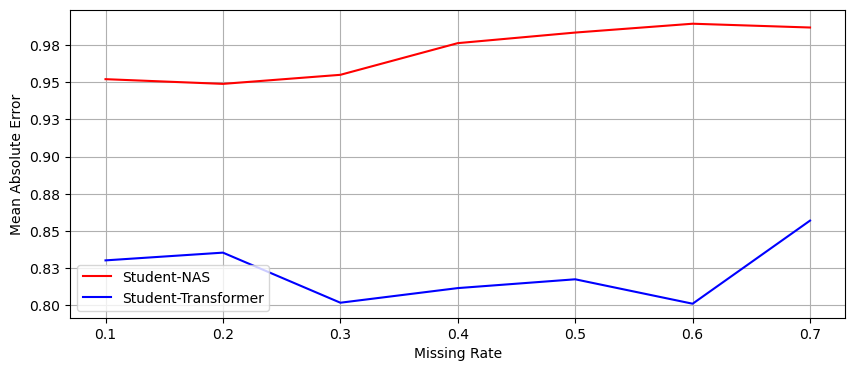}
		\label{iemocapfourmae}
	}
	\subfigure[IEMOCAP(six-class)]{
		\includegraphics[width=0.4\linewidth]{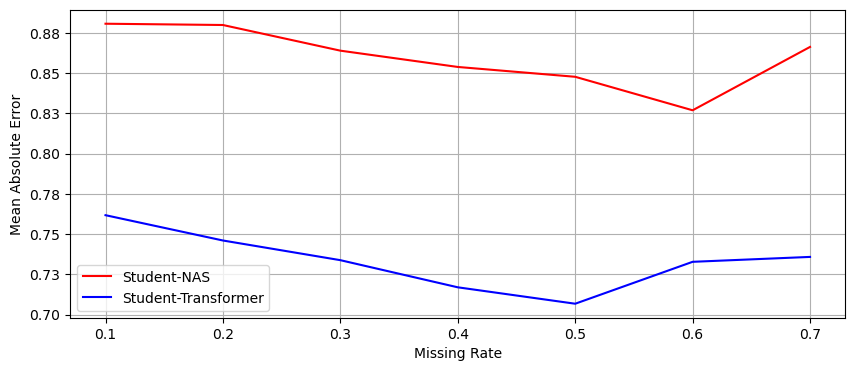}
		\label{iemocapsixmae}
	}
	\subfigure[CMU-MOSI]{
		\includegraphics[width=0.4\linewidth]{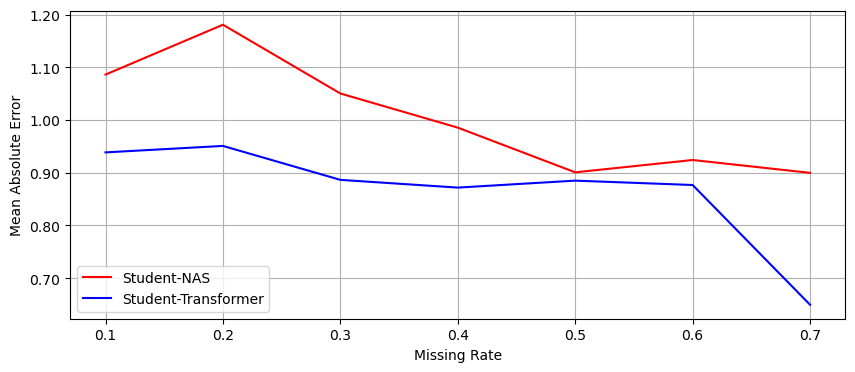}
		\label{mosimae}
	}
	\subfigure[CMU-MOSEI]{
		\includegraphics[width=0.4\linewidth]{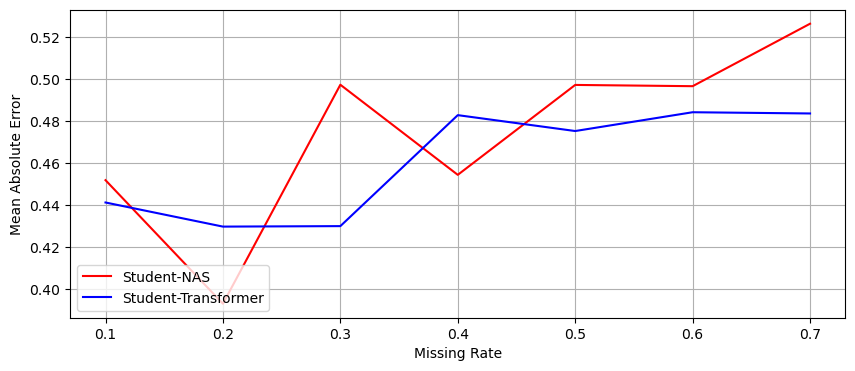}
		\label{moseimae}
	}
	
	\caption{The vector distance between $H_1^s$ and $H_1$ for different student models under the same teacher configuration.}
	\label{maeresults}
\end{figure*}
\begin{figure*}[!ht]
	\centering
	\subfigure[IEMOCAP(four-class)]{
		\includegraphics[width=0.48\linewidth]{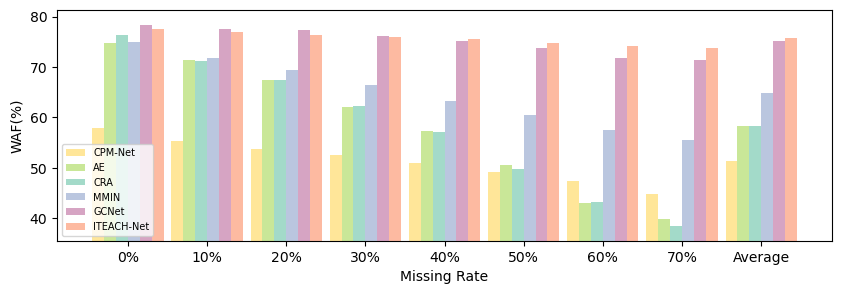}
		\label{iemocapfouriem}
	}
	\subfigure[IEMOCAP(six-class)]{
		\includegraphics[width=0.48\linewidth]{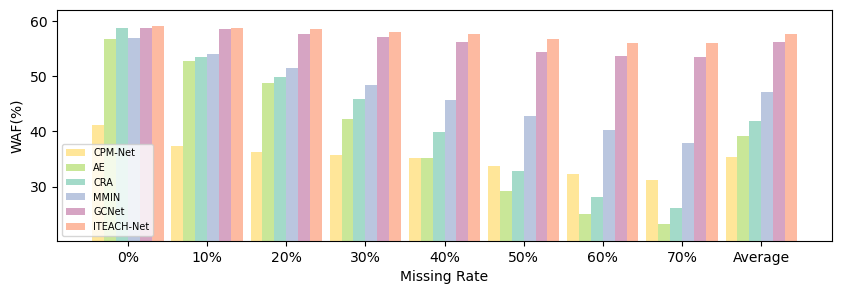}
		\label{iemocapsixiem}
	}
	\subfigure[CMU-MOSI]{
		\includegraphics[width=0.48\linewidth]{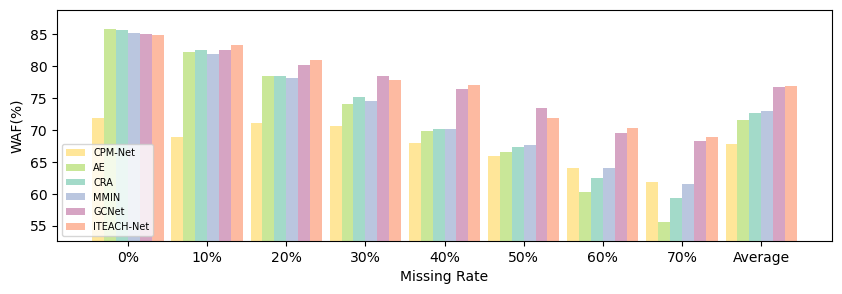}
		\label{mosiiem}
	}
	\subfigure[CMU-MOSEI]{
		\includegraphics[width=0.48\linewidth]{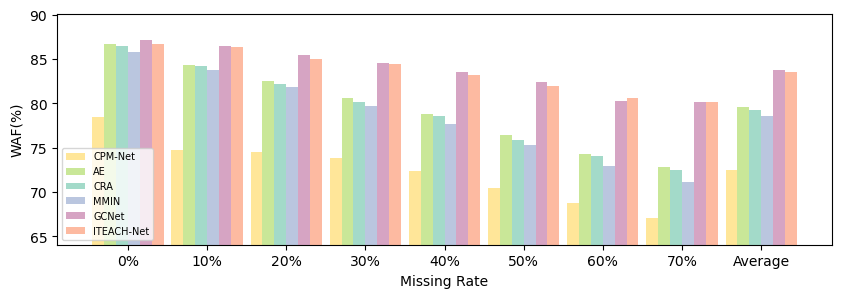}
		\label{moseiiem}
	}
	
	\caption{Comparison of classification performance under Independent Model Evaluation (IME). We present WAF scores (\%).}
	\label{IMresults}
\end{figure*}
\subsection{Distance Loss Performance}

Within the ITS framework, the student model primarily acquires knowledge from the teacher model through distance loss. To further investigate the benefits of distance loss on model robustness, we compare the vector mean absolute distance between the deep encodings $H^s_1$ of student models of varying complexities and the encoding $H_1$ of the teacher model. The results are presented in Figure \ref{maeresults}, where the red lines signifie the student model under the \textbf{NAS} setup, and the blue lines represent the student model in the \textbf{Transformer} configuration, with the x-axis denoting the data missing rate.

It is observable that, with the exception of the CMU-MOSEI task, the student models without NAS enhancement exhibit superior performance in predicting the hidden layer representations of the teacher model. However, while the NAS-enhanced student models show competitive reconstruction performance on the CMU-MOSEI task at data missing rates of 10\% to 40\%, they significantly underperform under conditions of extreme data incompleteness.


Integrating the findings with the results presented in Table \ref{ITSResults}, it is evident that a more robust student model performs poorly in reconstruction tasks. Therefore, we can conclude that while the knowledge imparted by the teacher model within the ITS framework is crucial, it only serves as a guide for the student model. In this guidance process, the student model can gradually learn the correct way to handle incomplete data suitable for itself, rather than merely focusing on imitating the teacher model. This also reveals the difference between the ITS method and imputation-based methods, where reconstruction performance does not represent the model's robustness.

\subsection{ITEACH-Net Performance}

%

\begin{figure*}[!h]
	\centering
	\subfigure[IEMOCAP(four-class)]{
		\includegraphics[width=0.45\linewidth]{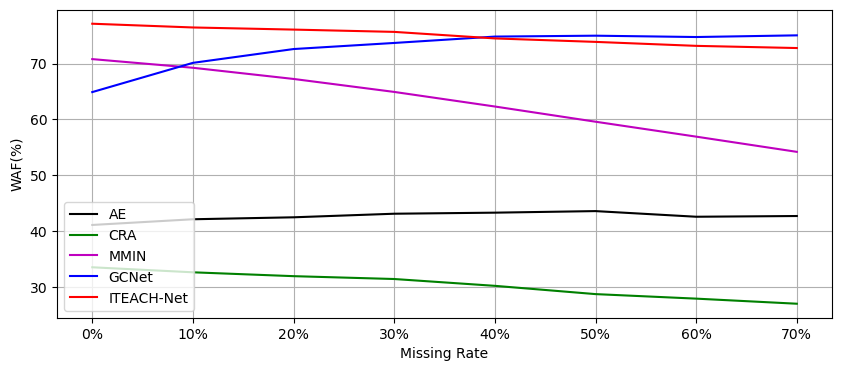}
		\label{iemocapfour70}
	}
	\subfigure[IEMOCAP(six-class)]{
		\includegraphics[width=0.45\linewidth]{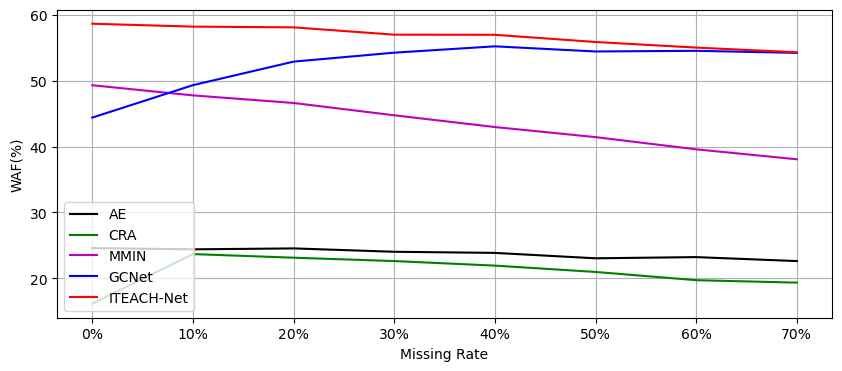}
		\label{iemocapsix70}
	}
	
	\caption{Comparison of classification performance at different missing rates for models trained with 70\% missing data. }
	\label{70results}
\end{figure*}

\subsubsection{Independent Model Evaluation Performance}
To demonstrate the outstanding performance of ITEACH-Net in ERC tasks, we maintained the same evaluation method as previous works in Figure \ref{IMresults}, enabling a fair comparison with the baselines.

Firstly, the Average metric reflect the overall performance of the model under different missing scenarios. ITEACH-Net surpasses the current state-of-the-art (sota) work, GCNet, by 0.52\%, 1.38\%, and 0.17\% in the four-class IEMOCAP, six-class IEMOCAP, and CMU-MOSI tasks, respectively, achieving the new sota performance. Although ITEACH-Net's performance in the CMU-MOSEI task is 0.19\% lower than GCNet, considering the carefully design of GCNet's network structure, this performance remains highly competitive.

Furthermore, observing ITEACH-Net's performance under different missing rates, we find that despite GCNet utilizing complex network structures to additionally process speaker information in conversation, there is still no significant difference between our method and GCNet across all four tasks. Even in severe missing scenarios, such as at a 70\% missing rate, our method demonstrates more pronounced advantages in the four-class IEMOCAP and six-class IEMOCAP tasks, achieving performance improvements of 2.46\% and 2.48\%, respectively. These results collectively demonstrate the excellent robustness of ITEACH-Net.

\begin{table}[!h]
	\centering
	\caption{The model parameters are implemented using the default settings provided in the official code of GCNet.}
	\label{Model Params}
	\begin{tabular}{c|c|c}
		\hline \hline
		Model                           & Module  & Parameters \\ \hline \hline
		AE\cite{bengio2006greedy}						& -			& 1.29M				\\
		CRA\cite{CRA}					& -			& 2.11M				\\
		MMIN\cite{MMIN}						& -			& 3.87M				\\
		GCNet\cite{GCNet}                           & -       & 34.14M      \\ \hline
		\multirow{2}{*}{ITEACH-Net}         & Teacher & 2.19M      \\
		& Student & 3.88M      \\ \hline \hline
	\end{tabular}
\end{table}

\subsubsection{Unified Model Evaluation Performance}

To more accurately demonstrate the model's robustness in dynamic missing rate scenarios, we further employ the UME method to evaluate their performance. As previous works did not provide experimental results for this evaluation method, all results presented here are reproduced by us. The parameters for each model during the reproduction process are listed in Table \ref{Model Params}. 

\begin{table*}[!h]
	\centering
	\caption{Comparison of classification performance under Random training strategy. The best performance is highlighted in bold.}
	\label{randomresults}
	\begin{tabular}{c|c|ccccccccc}
		\hline \hline
		\multirow{2}{*}{Dataset}                                                        & \multirow{2}{*}{Method} & \multicolumn{9}{c}{Missing Rate}                                                                                                                                                                                                                \\ \cline{3-11} 
		&                         & \multicolumn{1}{c|}{0.0$\uparrow$}   & \multicolumn{1}{c|}{0.1$\uparrow$}   & \multicolumn{1}{c|}{0.2$\uparrow$}   & \multicolumn{1}{c|}{0.3$\uparrow$}   & \multicolumn{1}{c|}{0.4$\uparrow$}   & \multicolumn{1}{c|}{0.5$\uparrow$}   & \multicolumn{1}{c|}{0.6$\uparrow$}   & \multicolumn{1}{c|}{0.7$\uparrow$}   & Average$\uparrow$ \\ \hline \hline
		\multirow{5}{*}{\begin{tabular}[c]{@{}c@{}}IEMOCAP\\ (four-class)\end{tabular}} & AE\cite{bengio2006greedy}                      & \multicolumn{1}{c|}{57.55} & \multicolumn{1}{c|}{56.43} & \multicolumn{1}{c|}{55.26} & \multicolumn{1}{c|}{53.59} & \multicolumn{1}{c|}{51.50} & \multicolumn{1}{c|}{49.53} & \multicolumn{1}{c|}{47.10} & \multicolumn{1}{c|}{45.16} & 52.02   \\
		& CRA\cite{CRA}                     & \multicolumn{1}{c|}{36.89} & \multicolumn{1}{c|}{35.76} & \multicolumn{1}{c|}{34.77} & \multicolumn{1}{c|}{33.45} & \multicolumn{1}{c|}{32.28} & \multicolumn{1}{c|}{30.72} & \multicolumn{1}{c|}{29.22} & \multicolumn{1}{c|}{28.07} & 32.65   \\
		& MMIN\cite{MMIN}                    & \multicolumn{1}{c|}{74.71} & \multicolumn{1}{c|}{72.88} & \multicolumn{1}{c|}{70.81} & \multicolumn{1}{c|}{68.15} & \multicolumn{1}{c|}{65.49} & \multicolumn{1}{c|}{62.22} & \multicolumn{1}{c|}{58.53} & \multicolumn{1}{c|}{55.85} & 66.08   \\
		& GCNet\cite{GCNet}                   & \multicolumn{1}{c|}{77.09} & \multicolumn{1}{c|}{77.00} & \multicolumn{1}{c|}{\textbf{76.47}} & \multicolumn{1}{c|}{\textbf{76.19}} & \multicolumn{1}{c|}{\textbf{75.30}} & \multicolumn{1}{c|}{\textbf{74.71}} & \multicolumn{1}{c|}{72.24} & \multicolumn{1}{c|}{69.94} & 74.87   \\
		& ITEACH-Net                  & \multicolumn{1}{c|}{\textbf{77.61}} & \multicolumn{1}{c|}{\textbf{77.23}} & \multicolumn{1}{c|}{76.46} & \multicolumn{1}{c|}{75.97} & \multicolumn{1}{c|}{75.12} & \multicolumn{1}{c|}{73.99} & \multicolumn{1}{c|}{\textbf{72.40}} & \multicolumn{1}{c|}{\textbf{71.73}} & \textbf{75.06}   \\ \hline
		\multirow{5}{*}{\begin{tabular}[c]{@{}c@{}}IEMOCAP\\ (six-class)\end{tabular}}  & AE\cite{bengio2006greedy}                      & \multicolumn{1}{c|}{28.10} & \multicolumn{1}{c|}{27.82} & \multicolumn{1}{c|}{27.51} & \multicolumn{1}{c|}{27.49} & \multicolumn{1}{c|}{27.00} & \multicolumn{1}{c|}{26.27} & \multicolumn{1}{c|}{25.78} & \multicolumn{1}{c|}{25.05} & 26.88   \\
		& CRA\cite{CRA}                     & \multicolumn{1}{c|}{24.15} & \multicolumn{1}{c|}{23.65} & \multicolumn{1}{c|}{23.04} & \multicolumn{1}{c|}{22.16} & \multicolumn{1}{c|}{21.39} & \multicolumn{1}{c|}{20.65} & \multicolumn{1}{c|}{19.38} & \multicolumn{1}{c|}{18.53} & 21.62   \\
		& MMIN\cite{MMIN}                    & \multicolumn{1}{c|}{57.00} & \multicolumn{1}{c|}{54.85} & \multicolumn{1}{c|}{52.79} & \multicolumn{1}{c|}{50.13} & \multicolumn{1}{c|}{47.33} & \multicolumn{1}{c|}{44.62} & \multicolumn{1}{c|}{41.24} & \multicolumn{1}{c|}{39.16} & 48.39   \\
		& GCNet\cite{GCNet}                   & \multicolumn{1}{c|}{56.05} & \multicolumn{1}{c|}{56.41} & \multicolumn{1}{c|}{56.18} & \multicolumn{1}{c|}{56.03} & \multicolumn{1}{c|}{55.87} & \multicolumn{1}{c|}{54.42} & \multicolumn{1}{c|}{52.93} & \multicolumn{1}{c|}{51.87} & 54.97   \\
		& ITEACH-Net                  & \multicolumn{1}{c|}{\textbf{59.90}} & \multicolumn{1}{c|}{\textbf{59.32}} & \multicolumn{1}{c|}{\textbf{58.61}} & \multicolumn{1}{c|}{\textbf{57.86}} & \multicolumn{1}{c|}{\textbf{57.01}} & \multicolumn{1}{c|}{\textbf{56.36}} & \multicolumn{1}{c|}{\textbf{55.29}} & \multicolumn{1}{c|}{\textbf{54.01}} & \textbf{57.30}   \\ \hline
		\multirow{5}{*}{CMUMOSI}                                                        & AE\cite{bengio2006greedy}                      & \multicolumn{1}{c|}{83.84} & \multicolumn{1}{c|}{79.80} & \multicolumn{1}{c|}{77.07} & \multicolumn{1}{c|}{70.78} & \multicolumn{1}{c|}{66.53} & \multicolumn{1}{c|}{63.67} & \multicolumn{1}{c|}{56.56} & \multicolumn{1}{c|}{51.91} & 68.77   \\
		& CRA\cite{CRA}                     & \multicolumn{1}{c|}{76.84} & \multicolumn{1}{c|}{73.87} & \multicolumn{1}{c|}{70.75} & \multicolumn{1}{c|}{66.82} & \multicolumn{1}{c|}{62.96} & \multicolumn{1}{c|}{58.93} & \multicolumn{1}{c|}{55.75} & \multicolumn{1}{c|}{50.90} & 64.60   \\
		& MMIN\cite{MMIN}                    & \multicolumn{1}{c|}{84.57} & \multicolumn{1}{c|}{80.74} & \multicolumn{1}{c|}{76.63} & \multicolumn{1}{c|}{74.69} & \multicolumn{1}{c|}{69.57} & \multicolumn{1}{c|}{64.28} & \multicolumn{1}{c|}{59.68} & \multicolumn{1}{c|}{57.05} & 70.90   \\
		& GCNet\cite{GCNet}                   & \multicolumn{1}{c|}{84.37} & \multicolumn{1}{c|}{82.62} & \multicolumn{1}{c|}{79.64} & \multicolumn{1}{c|}{77.86} & \multicolumn{1}{c|}{75.50} & \multicolumn{1}{c|}{73.12} & \multicolumn{1}{c|}{\textbf{71.34}} & \multicolumn{1}{c|}{67.11} & 76.44   \\
		& ITEACH-Net                  & \multicolumn{1}{c|}{\textbf{86.23}} & \multicolumn{1}{c|}{\textbf{84.22}} & \multicolumn{1}{c|}{\textbf{81.65}} & \multicolumn{1}{c|}{\textbf{79.27}} & \multicolumn{1}{c|}{\textbf{76.43}} & \multicolumn{1}{c|}{\textbf{75.18}} & \multicolumn{1}{c|}{70.65} & \multicolumn{1}{c|}{\textbf{68.96}} & \textbf{77.82}   \\ \hline
		\multirow{5}{*}{CMUMOSEI}                                                       & AE\cite{bengio2006greedy}                      & \multicolumn{1}{c|}{84.48} & \multicolumn{1}{c|}{82.82} & \multicolumn{1}{c|}{81.20} & \multicolumn{1}{c|}{79.13} & \multicolumn{1}{c|}{77.33} & \multicolumn{1}{c|}{75.49} & \multicolumn{1}{c|}{73.13} & \multicolumn{1}{c|}{71.24} & 78.10   \\
		& CRA\cite{CRA}                     & \multicolumn{1}{c|}{75.48} & \multicolumn{1}{c|}{74.06} & \multicolumn{1}{c|}{72.56} & \multicolumn{1}{c|}{70.95} & \multicolumn{1}{c|}{69.18} & \multicolumn{1}{c|}{67.24} & \multicolumn{1}{c|}{64.84} & \multicolumn{1}{c|}{63.56} & 69.73   \\
		& MMIN\cite{MMIN}                    & \multicolumn{1}{c|}{85.75} & \multicolumn{1}{c|}{83.12} & \multicolumn{1}{c|}{80.55} & \multicolumn{1}{c|}{77.52} & \multicolumn{1}{c|}{75.10} & \multicolumn{1}{c|}{71.44} & \multicolumn{1}{c|}{68.17} & \multicolumn{1}{c|}{65.38} & 75.88   \\
		& GCNet\cite{GCNet}                   & \multicolumn{1}{c|}{86.84} & \multicolumn{1}{c|}{86.06} & \multicolumn{1}{c|}{\textbf{85.25}} & \multicolumn{1}{c|}{84.48} & \multicolumn{1}{c|}{\textbf{83.80}} & \multicolumn{1}{c|}{82.13} & \multicolumn{1}{c|}{80.66} & \multicolumn{1}{c|}{79.64} & 83.61   \\
		& ITEACH-Net                  & \multicolumn{1}{c|}{\textbf{87.40}} & \multicolumn{1}{c|}{\textbf{86.50}} & \multicolumn{1}{c|}{85.11} & \multicolumn{1}{c|}{\textbf{84.59}} & \multicolumn{1}{c|}{82.81} & \multicolumn{1}{c|}{\textbf{82.30}} & \multicolumn{1}{c|}{\textbf{80.66}} & \multicolumn{1}{c|}{\textbf{79.79}} & \textbf{83.65}   \\ \hline \hline
		
	\end{tabular}
\end{table*}

To demonstrate the limitations of the IME, we present in Figure \ref{70results} the performance of different methods at various missing rates in the four-class and six-class IEMOCAP tasks when the model is trained only on data with a 70\% missing rate. The results show that AE and CRA almost completely lose their predictive ability. Although GCNet can perform well on incomplete data, due to the limitations of the training framework, it cannot handle complete data effectively, leading to a decline in model performance as the missing rate decreases. The MMIN method based on distillation maintains the ability to process complete data, but its performance drops significantly when dealing with incomplete data. In contrast, ITEACH-Net learns to handle complete data effectively, not only improving performance as the missing rate decreases but also maintaining stable performance across different missing rates. These results aptly reflect the necessity of testing model robustness using UME.

However, the optimal training approach to fully leverage the potential of the model within the UME method remains an open question. In this paper, we compare the performance of models under two different training strategies:

\textbf{Random Strategy:}
To equip the model with the capability to process data with varying degrees of incompleteness, a straightforward approach involves blending training data with different missing rates. Under this training strategy, each batch of training data sampled by the model is randomly assigned a missing rate ranging from 0\% to 70\%, ensuring that each batch experiences a distinct level of incompleteness.

As illustrated in Table \ref{randomresults}, it is observable that ITEACH-Net achieves optimal results across the Average metrics of four tasks, particularly demonstrating a significant performance advantage in the six-class IEMOCAP and CMU-MOSI tasks. Similarly, ITEACH-Net consistently outperformed all baselines across various missing rates in these two tasks, while in the four-class IEMOCAP and CMU-MOSEI tasks, it alternately excels or is comparable to GCNet.

\begin{table*}[!h]
	\centering
	\caption{Comparison of classification performance under Progressive training strategy. The best performance is highlighted in bold.}
	\label{proresults}
	\begin{tabular}{c|c|ccccccccc}
		\hline \hline
		\multirow{2}{*}{Dataset}                                                        & \multirow{2}{*}{Method} & \multicolumn{9}{c}{Missing Rate}                                                                                                                                                                                                                \\ \cline{3-11} 
		&                         & \multicolumn{1}{c|}{0.0$\uparrow$}   & \multicolumn{1}{c|}{0.1$\uparrow$}   & \multicolumn{1}{c|}{0.2$\uparrow$}   & \multicolumn{1}{c|}{0.3$\uparrow$}   & \multicolumn{1}{c|}{0.4$\uparrow$}   & \multicolumn{1}{c|}{0.5$\uparrow$}   & \multicolumn{1}{c|}{0.6$\uparrow$}   & \multicolumn{1}{c|}{0.7$\uparrow$}   & Average$\uparrow$ \\ \hline \hline
		\multirow{5}{*}{\begin{tabular}[c]{@{}c@{}}IEMOCAP\\ (four-class)\end{tabular}} & AE\cite{bengio2006greedy}                      & \multicolumn{1}{c|}{67.60} & \multicolumn{1}{c|}{66.23} & \multicolumn{1}{c|}{64.09} & \multicolumn{1}{c|}{62.28} & \multicolumn{1}{c|}{59.91} & \multicolumn{1}{c|}{56.97} & \multicolumn{1}{c|}{53.73} & \multicolumn{1}{c|}{51.41} & 60.28   \\
		& CRA\cite{CRA}                     & \multicolumn{1}{c|}{61.34} & \multicolumn{1}{c|}{59.12} & \multicolumn{1}{c|}{56.59} & \multicolumn{1}{c|}{53.90} & \multicolumn{1}{c|}{51.12} & \multicolumn{1}{c|}{48.10} & \multicolumn{1}{c|}{43.98} & \multicolumn{1}{c|}{41.51} & 51.96   \\
		& MMIN\cite{MMIN}                    & \multicolumn{1}{c|}{74.77} & \multicolumn{1}{c|}{72.74} & \multicolumn{1}{c|}{70.34} & \multicolumn{1}{c|}{67.59} & \multicolumn{1}{c|}{64.98} & \multicolumn{1}{c|}{61.81} & \multicolumn{1}{c|}{58.30} & \multicolumn{1}{c|}{56.05} & 65.82   \\
		& GCNet\cite{GCNet}                   & \multicolumn{1}{c|}{71.67} & \multicolumn{1}{c|}{73.53} & \multicolumn{1}{c|}{74.90} & \multicolumn{1}{c|}{\textbf{75.96}} & \multicolumn{1}{c|}{\textbf{75.72}} & \multicolumn{1}{c|}{\textbf{76.20}} & \multicolumn{1}{c|}{\textbf{75.48}} & \multicolumn{1}{c|}{\textbf{74.80}} & 74.78   \\
		& ITEACH-Net                  & \multicolumn{1}{c|}{\textbf{77.47}} & \multicolumn{1}{c|}{\textbf{76.86}} & \multicolumn{1}{c|}{\textbf{76.78}} & \multicolumn{1}{c|}{75.86} & \multicolumn{1}{c|}{75.26} & \multicolumn{1}{c|}{74.33} & \multicolumn{1}{c|}{73.73} & \multicolumn{1}{c|}{71.79} & \textbf{75.26}   \\ \hline
		\multirow{5}{*}{\begin{tabular}[c]{@{}c@{}}IEMOCAP\\ (six-class)\end{tabular}}  & AE\cite{bengio2006greedy}                      & \multicolumn{1}{c|}{48.08} & \multicolumn{1}{c|}{46.38} & \multicolumn{1}{c|}{44.71} & \multicolumn{1}{c|}{43.20} & \multicolumn{1}{c|}{41.08} & \multicolumn{1}{c|}{39.07} & \multicolumn{1}{c|}{36.37} & \multicolumn{1}{c|}{34.28} & 41.64   \\
		& CRA\cite{CRA}                     & \multicolumn{1}{c|}{36.01} & \multicolumn{1}{c|}{34.95} & \multicolumn{1}{c|}{34.07} & \multicolumn{1}{c|}{32.60} & \multicolumn{1}{c|}{31.17} & \multicolumn{1}{c|}{29.68} & \multicolumn{1}{c|}{26.51} & \multicolumn{1}{c|}{25.24} & 31.28   \\
		& MMIN\cite{MMIN}                    & \multicolumn{1}{c|}{57.16} & \multicolumn{1}{c|}{54.81} & \multicolumn{1}{c|}{53.05} & \multicolumn{1}{c|}{50.15} & \multicolumn{1}{c|}{47.85} & \multicolumn{1}{c|}{44.79} & \multicolumn{1}{c|}{41.39} & \multicolumn{1}{c|}{38.90} & 48.51   \\
		& GCNet\cite{GCNet}                   & \multicolumn{1}{c|}{51.88} & \multicolumn{1}{c|}{53.88} & \multicolumn{1}{c|}{55.34} & \multicolumn{1}{c|}{55.53} & \multicolumn{1}{c|}{55.91} & \multicolumn{1}{c|}{56.22} & \multicolumn{1}{c|}{\textbf{56.16}} & \multicolumn{1}{c|}{\textbf{55.27}} & 55.03   \\
		& ITEACH-Net                  & \multicolumn{1}{c|}{\textbf{59.09}} & \multicolumn{1}{c|}{\textbf{58.72}} & \multicolumn{1}{c|}{\textbf{58.13}} & \multicolumn{1}{c|}{\textbf{57.53}} & \multicolumn{1}{c|}{\textbf{57.18}} & \multicolumn{1}{c|}{\textbf{56.39}} & \multicolumn{1}{c|}{55.53} & \multicolumn{1}{c|}{55.06} & \textbf{57.21}  \\ \hline
		\multirow{5}{*}{CMUMOSI}                                                        & AE\cite{bengio2006greedy}                      & \multicolumn{1}{c|}{\textbf{85.66}} & \multicolumn{1}{c|}{81.72} & \multicolumn{1}{c|}{78.29} & \multicolumn{1}{c|}{74.14} & \multicolumn{1}{c|}{69.46} & \multicolumn{1}{c|}{62.81} & \multicolumn{1}{c|}{58.13} & \multicolumn{1}{c|}{54.23} & 70.55   \\
		& CRA\cite{CRA}                     & \multicolumn{1}{c|}{85.00} & \multicolumn{1}{c|}{80.90} & \multicolumn{1}{c|}{77.01} & \multicolumn{1}{c|}{72.03} & \multicolumn{1}{c|}{67.46} & \multicolumn{1}{c|}{61.86} & \multicolumn{1}{c|}{56.75} & \multicolumn{1}{c|}{52.24} & 69.16   \\
		& MMIN\cite{MMIN}                    & \multicolumn{1}{c|}{84.62} & \multicolumn{1}{c|}{79.84} & \multicolumn{1}{c|}{76.32} & \multicolumn{1}{c|}{71.39} & \multicolumn{1}{c|}{68.05} & \multicolumn{1}{c|}{61.95} & \multicolumn{1}{c|}{56.40} & \multicolumn{1}{c|}{52.93} & 68.94   \\
		& GCNet\cite{GCNet}                   & \multicolumn{1}{c|}{68.78} & \multicolumn{1}{c|}{70.91} & \multicolumn{1}{c|}{70.72} & \multicolumn{1}{c|}{70.11} & \multicolumn{1}{c|}{69.45} & \multicolumn{1}{c|}{69.72} & \multicolumn{1}{c|}{68.28} & \multicolumn{1}{c|}{\textbf{68.16}} & 69.52   \\
		& ITEACH-Net                  & \multicolumn{1}{c|}{84.40} & \multicolumn{1}{c|}{\textbf{82.39}} & \multicolumn{1}{c|}{\textbf{80.54}} & \multicolumn{1}{c|}{\textbf{78.86}} & \multicolumn{1}{c|}{\textbf{75.62}} & \multicolumn{1}{c|}{\textbf{74.30}} & \multicolumn{1}{c|}{\textbf{69.92}} & \multicolumn{1}{c|}{65.77} & \textbf{76.48}   \\ \hline
		\multirow{5}{*}{CMUMOSEI}                                                       & AE\cite{bengio2006greedy}                      & \multicolumn{1}{c|}{85.19} & \multicolumn{1}{c|}{83.14} & \multicolumn{1}{c|}{81.50} & \multicolumn{1}{c|}{79.33} & \multicolumn{1}{c|}{77.00} & \multicolumn{1}{c|}{74.43} & \multicolumn{1}{c|}{71.52} & \multicolumn{1}{c|}{70.19} & 77.79   \\
		& CRA\cite{CRA}                     & \multicolumn{1}{c|}{85.73} & \multicolumn{1}{c|}{83.18} & \multicolumn{1}{c|}{80.61} & \multicolumn{1}{c|}{77.51} & \multicolumn{1}{c|}{74.69} & \multicolumn{1}{c|}{71.94} & \multicolumn{1}{c|}{67.84} & \multicolumn{1}{c|}{65.49} & 75.88   \\
		& MMIN\cite{MMIN}                    & \multicolumn{1}{c|}{85.21} & \multicolumn{1}{c|}{82.83} & \multicolumn{1}{c|}{80.19} & \multicolumn{1}{c|}{78.09} & \multicolumn{1}{c|}{74.53} & \multicolumn{1}{c|}{71.53} & \multicolumn{1}{c|}{67.90} & \multicolumn{1}{c|}{66.08} & 75.79   \\
		& GCNet\cite{GCNet}                   & \multicolumn{1}{c|}{82.81} & \multicolumn{1}{c|}{82.67} & \multicolumn{1}{c|}{82.72} & \multicolumn{1}{c|}{82.31} & \multicolumn{1}{c|}{82.03} & \multicolumn{1}{c|}{\textbf{81.81}} & \multicolumn{1}{c|}{\textbf{81.12}} & \multicolumn{1}{c|}{\textbf{79.99}} & 81.93   \\
		& ITEACH-Net                  & \multicolumn{1}{c|}{\textbf{86.91}} & \multicolumn{1}{c|}{\textbf{85.78}} & \multicolumn{1}{c|}{\textbf{84.86}} & \multicolumn{1}{c|}{\textbf{84.16}} & \multicolumn{1}{c|}{\textbf{82.77}} & \multicolumn{1}{c|}{81.60} & \multicolumn{1}{c|}{79.80} & \multicolumn{1}{c|}{78.91} & \textbf{83.10}   \\ \hline\hline
		
	\end{tabular}
\end{table*}

\textbf{Progressive Strategy:}
Compared to handling incomplete data, classifying tasks with complete data is evidently simpler. Hence, we aim to familiarize the model with complete data modeling initially, gradually introducing it to incomplete data in the hope that it can leverage previously acquired knowledge for better handling of incomplete data. Specifically, we start with a 0\% missing rate and increase it of the training data by 10\% every 10 epochs, up to a maximum of 70\%. This progressive strategy allows the model to adapt to incomplete data more smoothly.

As depicted in Table \ref{proresults}, ITEACH-Net consistently achieved optimal results on the Average metrics across all four tasks, with a particularly distinct performance advantage in all tasks except the four-class IEMOCAP. Although GCNet outperforms ITEACH-Net in the four-class IEMOCAP, six-class IEMOCAP and CMU-MOSEI tasks under severe data missing scenarios, it fails to maintain performance at lower missing rates, highlighting a drawback of the previous training frameworks. ITEACH-Net effectively mitigates this issue.


Furthermore, examining the performance of different methods across various missing rates, baselines under the Random training strategy show a balanced performance across different data missing rate scenarios but rarely achieve standout results. 
With the Progressive training strategy, baselines tend to exhibit a biased proficiency, performing well at a certain missing rate but declining as the missing rate changes, whether it increases or decreases. This indicates that the model is unable to learn how to process data at different missing rates simultaneously.
However, regardless of the training strategy employed, ITEACH-Net consistently demonstrates a balanced performance, exhibiting exceptional capability in processing both incomplete and complete data.

\subsection{Visualization of Embedding Space}

\begin{figure*}[!h]
	\centering
	\subfigure[Teacher]{
		\includegraphics[width=0.21\linewidth]{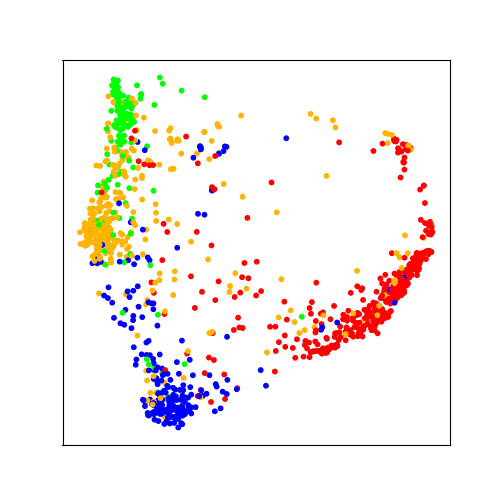}
		\label{IEMOCAPFour_teacher}
	}
	\hspace{-0.65cm}
	\subfigure[Student-0.0]{
		\includegraphics[width=0.21\linewidth]{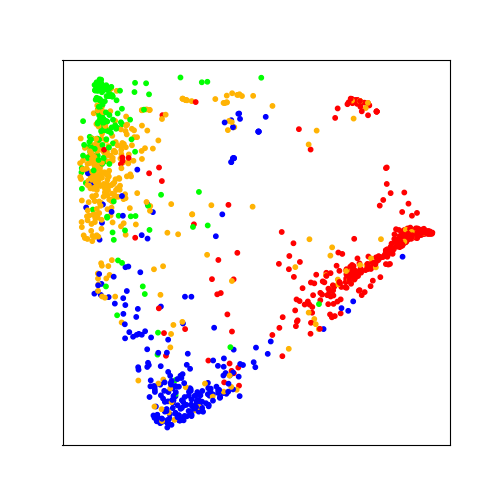}
		\label{IEMOCAPFour_student_0}
	}
	\hspace{-0.65cm}
	\subfigure[Student-0.1]{
		\includegraphics[width=0.21\linewidth]{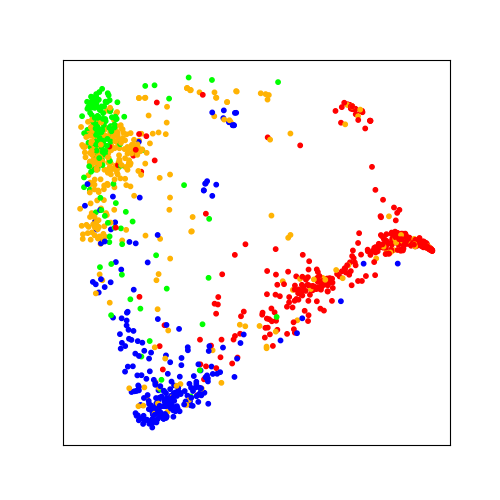}
		\label{IEMOCAPFour_student_1}
	}
	\hspace{-0.65cm}
	\subfigure[Student-0.2]{
		\includegraphics[width=0.21\linewidth]{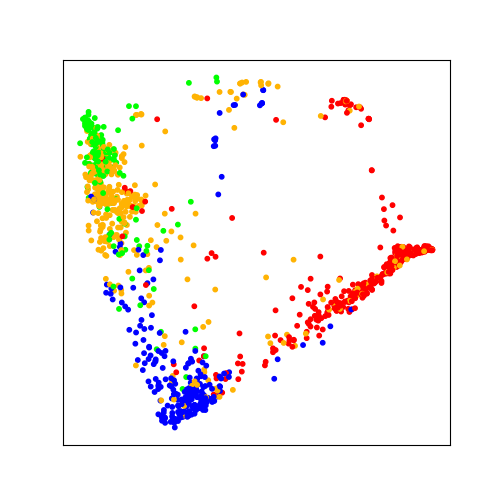}
		\label{IEMOCAPFour_student_2}
	}
	\hspace{-0.65cm}
	\subfigure[Student-0.3]{
		\includegraphics[width=0.21\linewidth]{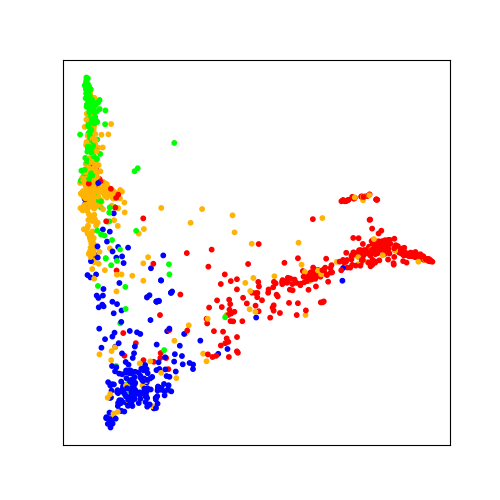}
		\label{IEMOCAPFour_student_3}
	}
	\hspace{-0.65cm}
	\subfigure[Student-0.4]{
		\includegraphics[width=0.21\linewidth]{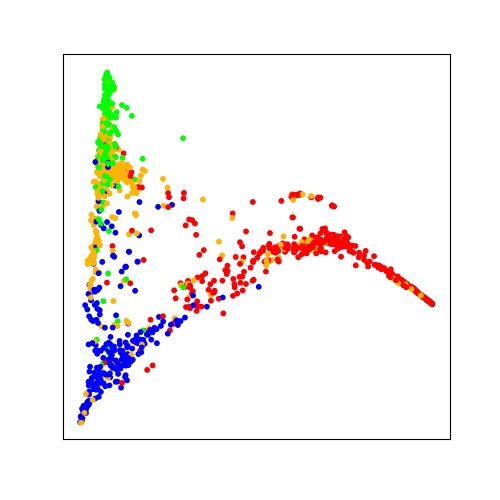}
		\label{IEMOCAPFour_student_4}
	}
	\hspace{-0.65cm}
	\subfigure[Student-0.5]{
		\includegraphics[width=0.21\linewidth]{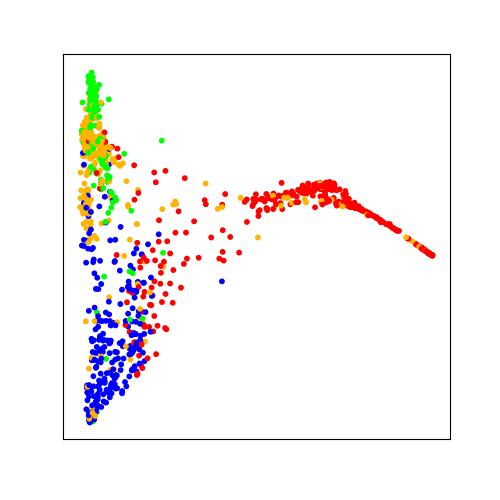}
		\label{IEMOCAPFour_student_5}
	}
	\hspace{-0.65cm}
	\subfigure[Student-0.6]{
		\includegraphics[width=0.21\linewidth]{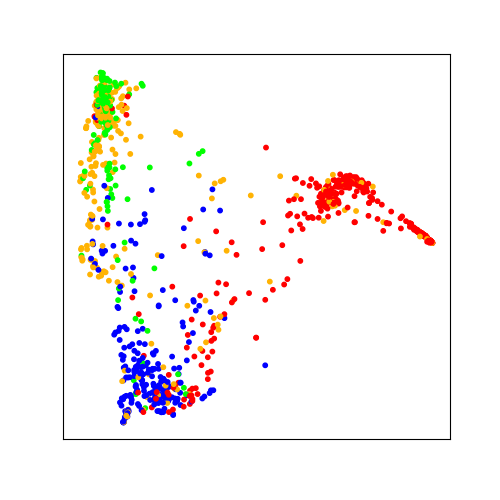}
		\label{IEMOCAPFour_student_6}
	}
	\hspace{-0.65cm}
	\subfigure[Student-0.7]{
		\includegraphics[width=0.21\linewidth]{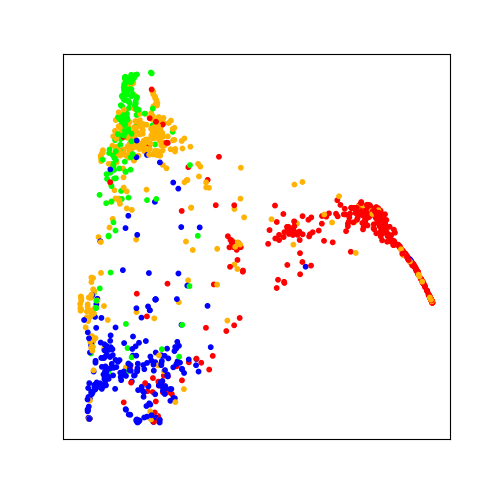}
		\label{IEMOCAPFour_student_7}
	}

	\caption{t-SNE visualization results on the four-class IEMOCAP test set with the increasing data missing rates. Student-0.1 represents the results achieved by the Student model when the data missing rate is 10\%. We use red, blue, yellow, and green to represent happiness, sadness, neutral, and anger, respectively.}
	\label{Embedding Space}
\end{figure*}

To qualitatively analyze the robustness of ITEACH-Net in the face of dynamic data missing rates, we visualized the latent representations under different missing scenarios on the four-class IEMOCAP test set. Figure \ref{Embedding Space} presents the t-SNE \cite{van2008visualizing} visualization results. Subfigure \ref{IEMOCAPFour_teacher} represents the encoding results of the Teacher model using complete data, while subfigures \ref{IEMOCAPFour_student_0} to \ref{IEMOCAPFour_student_7} depict the encoding results of the Student model under various data missing rates. It is observable that, in dynamic missing scenarios, the Student model still effectively emulates the knowledge of the Teacher during the encoding complete data. Moreover, as the data missing rate increases, although the variance of the same class data representation gradually enlarges, the overall distribution pattern remains unchanged, and there is still a clear distance between different class data representations. The stable data embedding space distribution demonstrates the robustness of ITEACH-Net in handling different missing scenarios.

\subsection{Visualization of Confusion Matrix}

\begin{figure*}[!h]
	\centering
	\subfigure[Student-0.0]{
		\includegraphics[width=0.242\linewidth]{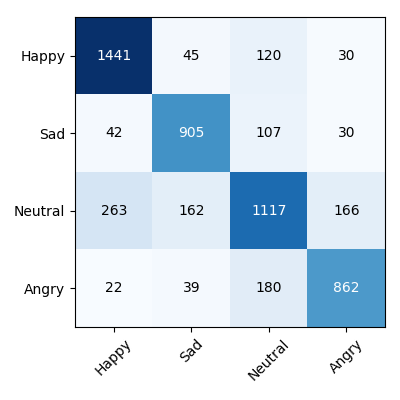}
		\label{IEMOCAPFour_cf_student_0}
	}
	\hspace{-0.45cm}
	\subfigure[Student-0.2]{
		\includegraphics[width=0.24\linewidth]{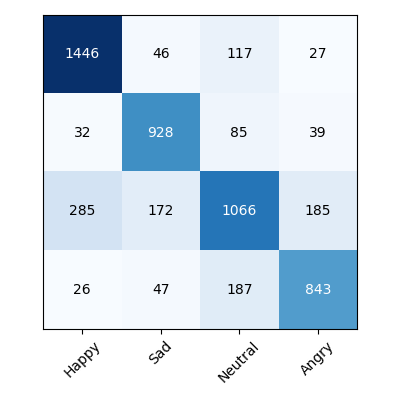}
		\label{IEMOCAPFour_cf_student_2}
	}
	\hspace{-0.45cm}
	\subfigure[Student-0.4]{
		\includegraphics[width=0.24\linewidth]{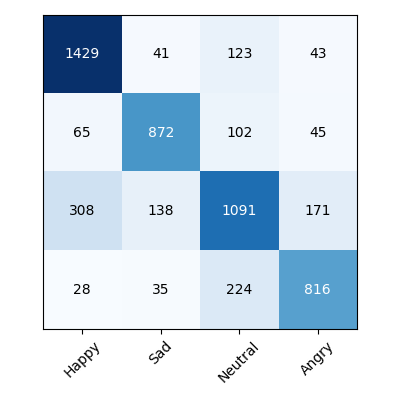}
		\label{IEMOCAPFour_cf_student_4}
	}
	\hspace{-0.45cm}
	\subfigure[Student-0.7]{
		\includegraphics[width=0.24\linewidth]{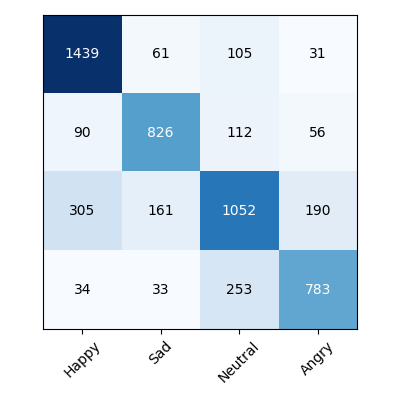}
		
	}

	\caption{Visualization results of confusion matrices on the four-class IEMOCAP with the increasing data missing rates.}
	\label{Confusion Matrixs}
\end{figure*}

To delve into the processing results of ITEACH-Net on imbalanced category data under different missing rates, we further visualized the confusion matrix of the model's prediction results in the Four-class IEMOCAP. The results are shown in Figure \ref{Confusion Matrixs}. It can be seen that the prediction accuracy for the Sad, Neutral, and Angry categories decreases with the increase in data missing rates. Among them, the Neutral category has the largest number of samples, totaling 1708, while the Angry category has the least, with only 1103 samples. In contrast, the Happy category, which has a larger number of samples, maintains a relatively stable prediction accuracy under different missing rates. Based on these results, it can be concluded that ITEACH-Net performs stably in the relatively simple task of Happy emotion detection in dynamic missing scenarios, and does not suffer from model collapse due to data imbalance in other emotion detection tasks. Similar prediction distributions under various missing rates also demonstrate the stability of ITEACH-Net in handling different missing scenarios.

\section{Conclusion}

In this paper, we propose a novel framework, ITEACH-Net, for incomplete multimodal learning in ERC. ITEACH-Net has a new ECCE module to capture emotional context changes, and experimental results verify the significance of this unique pattern in ERC. It also includes a novel ITS training framework, allowing a complex student model to leverage incomplete data, following the performance of a simple teacher model trained on complete data. This design provides the student model with sufficient capability to handle incomplete data, and experimental results demonstrate the importance of a complex student model in incomplete learning. Finally, using an evaluation method, we demonstrate that previous approaches struggle to effectively handle varying data missing rate scenarios when dealing with incomplete data. Concurrently, experimental results showcase the superiority of ITEACH-Net in incomplete learning.

In the future, we aim to expand ITEACH-Net to focus on real-time analysis scenarios, considering how to determine the emotional context patterns of the current time node based on data from previous time nodes. Additionally, beyond emotion recognition in conversation, we will also utilize ITEACH-Net to address the issue of modality missing in other types of conversation understanding tasks.

\section{Acknowledgements}
This work is supported by the National Natural Science Foundation of China (NSFC) ( No.62276259, No.62201572, No.U21B2010, No.62271083, No.62306316)



\bibliographystyle{elsarticle-num} 

\bibliography{airefdata}

%
%
%

\newpage

\appendix


\section{Embedding Space}

We further provide t-SNE visualization results on other three tasks. Stable embedding distributions at different missing rates demonstrate the model's stability and robustness when dealing with incomplete data.

\begin{figure*}[b!h]
	\centering
	\subfigure[Teacher]{
		\includegraphics[width=0.21\linewidth]{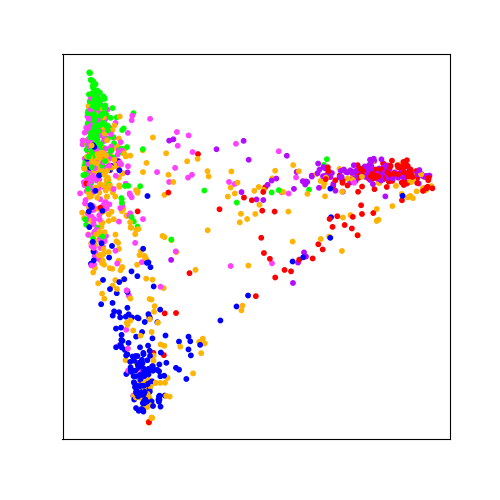}
		\label{IEMOCAPSix_teacher}
	}
	\hspace{-0.65cm}
	\subfigure[Student-0.0]{
		\includegraphics[width=0.21\linewidth]{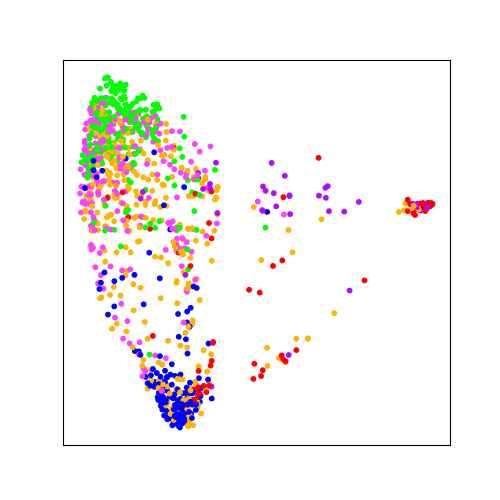}
		\label{IEMOCAPSix_student_0}
	}
	\hspace{-0.65cm}
	\subfigure[Student-0.1]{
		\includegraphics[width=0.21\linewidth]{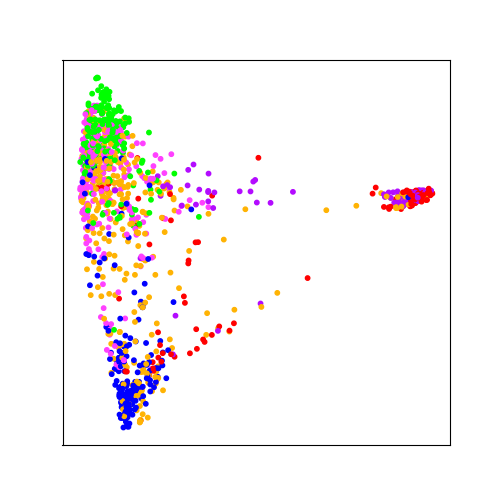}
		\label{IEMOCAPSix_student_1}
	}
	\hspace{-0.65cm}
	\subfigure[Student-0.2]{
		\includegraphics[width=0.21\linewidth]{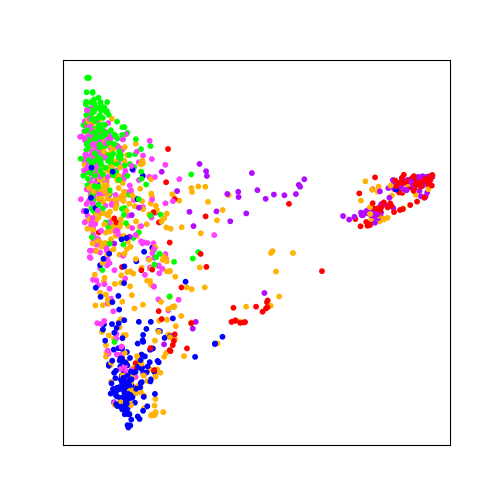}
		\label{IEMOCAPSix_student_2}
	}
	\hspace{-0.65cm}
	\subfigure[Student-0.3]{
		\includegraphics[width=0.21\linewidth]{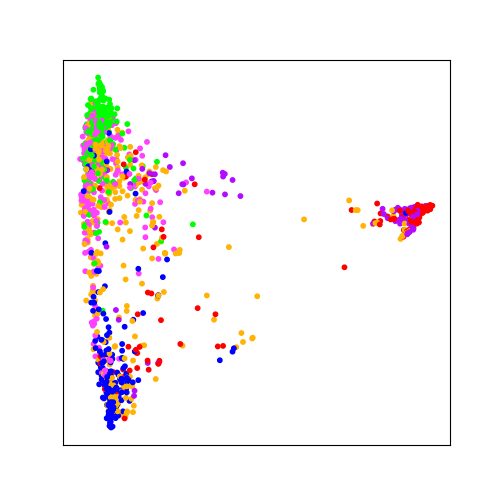}
		\label{IEMOCAPSix_student_3}
	}
	\hspace{-0.65cm}
	\subfigure[Student-0.4]{
		\includegraphics[width=0.21\linewidth]{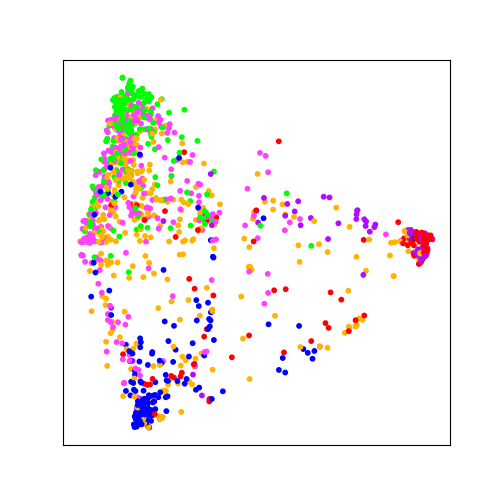}
		\label{IEMOCAPSix_student_4}
	}
	\hspace{-0.65cm}
	\subfigure[Student-0.5]{
		\includegraphics[width=0.21\linewidth]{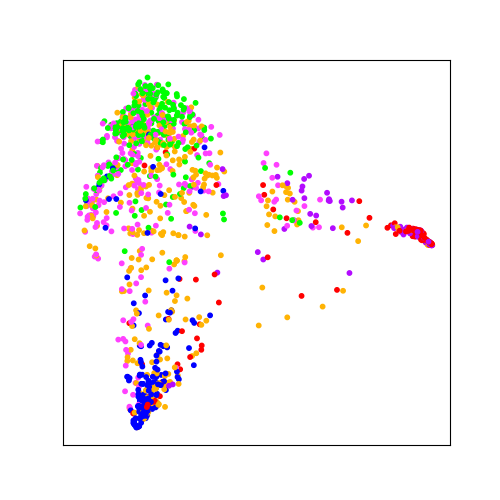}
		\label{IEMOCAPSix_student_5}
	}
	\hspace{-0.65cm}
	\subfigure[Student-0.6]{
		\includegraphics[width=0.21\linewidth]{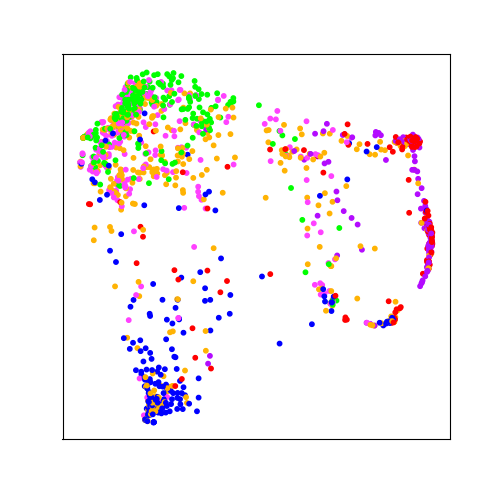}
		\label{IEMOCAPSix_student_6}
	}
	\hspace{-0.65cm}
	\subfigure[Student-0.7]{
		\includegraphics[width=0.21\linewidth]{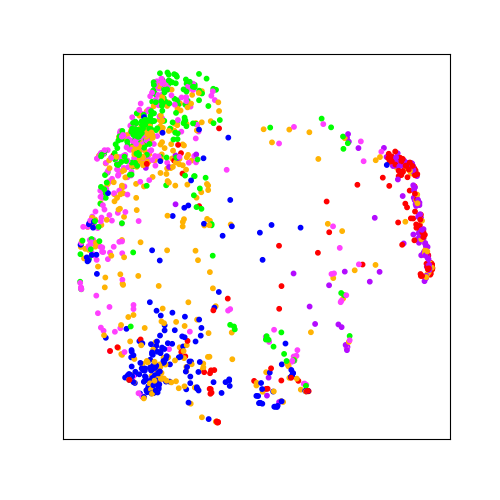}
		\label{IEMOCAPSix_student_7}
	}

	\caption{t-SNE visualization results on the Six-class IEMOCAP test set with the increasing data missing rates. Student-0.1 represents the results achieved by the Student model when the data missing rate is 10\%. We use red, blue, yellow, green, purple and pink to represent happiness, sadness, neutral, anger, excited and frustrated respectively.}
	\label{Embedding Space Six}
\end{figure*}
\begin{figure*}[ht]
	\centering
	\subfigure[Teacher]{
		\includegraphics[width=0.21\linewidth]{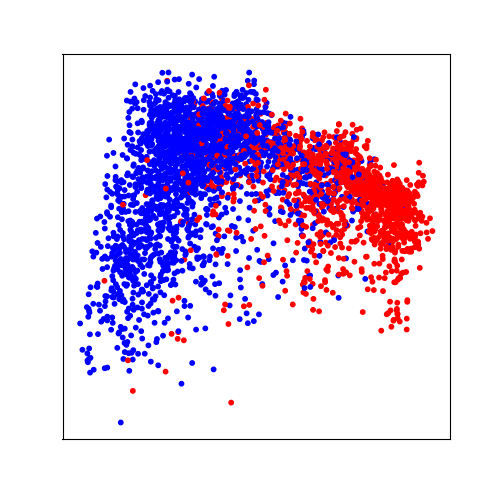}
		\label{MOSEI_teacher}
	}
	\hspace{-0.65cm}
	\subfigure[Student-0.0]{
		\includegraphics[width=0.21\linewidth]{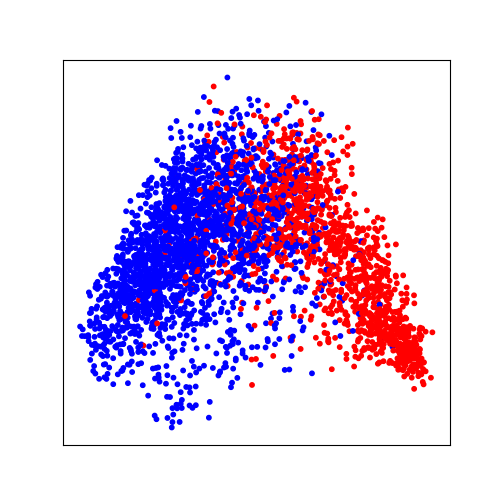}
		\label{MOSEI_student_0}
	}
	\hspace{-0.65cm}
	\subfigure[Student-0.1]{
		\includegraphics[width=0.21\linewidth]{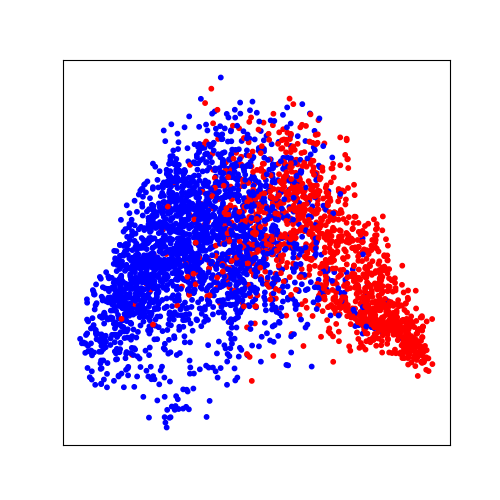}
		\label{MOSEI_student_1}
	}
	\hspace{-0.65cm}
	\subfigure[Student-0.2]{
		\includegraphics[width=0.21\linewidth]{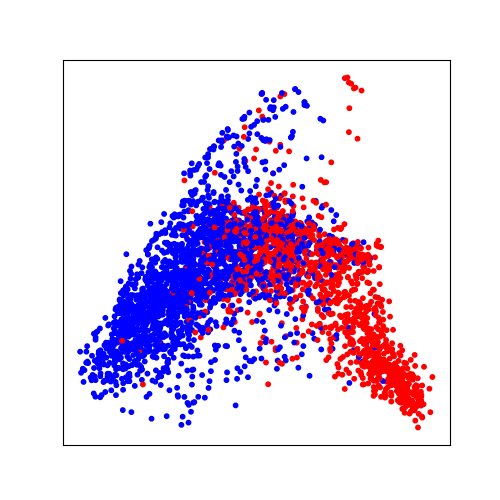}
		\label{MOSEI_student_2}
	}
	\hspace{-0.65cm}
	\subfigure[Student-0.3]{
		\includegraphics[width=0.21\linewidth]{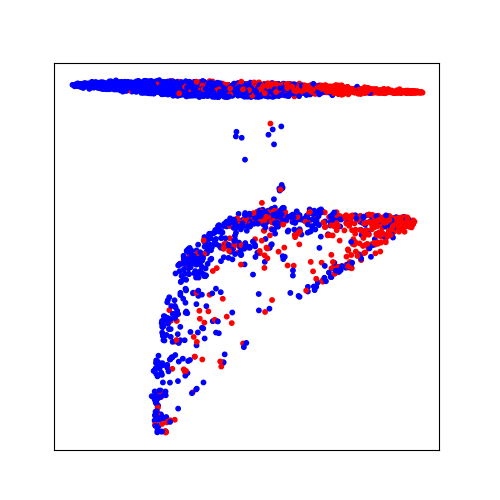}
		\label{MOSEI_student_3}
	}
	\hspace{-0.65cm}
	\subfigure[Student-0.4]{
		\includegraphics[width=0.21\linewidth]{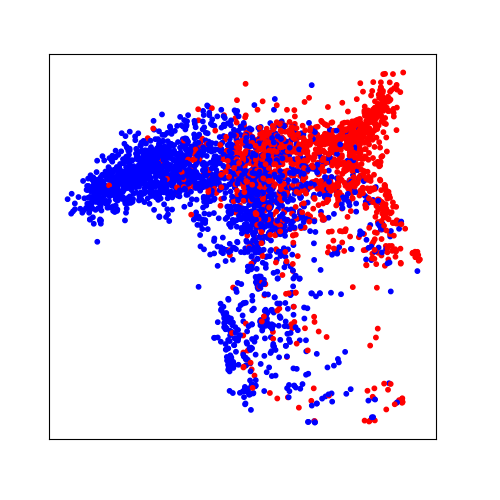}
		\label{MOSEI_student_4}
	}
	\hspace{-0.65cm}
	\subfigure[Student-0.5]{
		\includegraphics[width=0.21\linewidth]{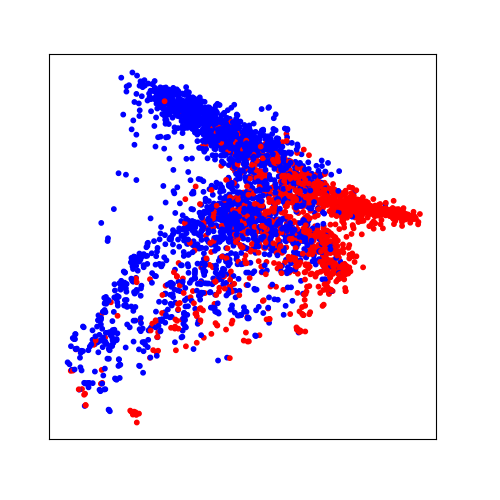}
		\label{MOSEI_student_5}
	}
	\hspace{-0.65cm}
	\subfigure[Student-0.6]{
		\includegraphics[width=0.21\linewidth]{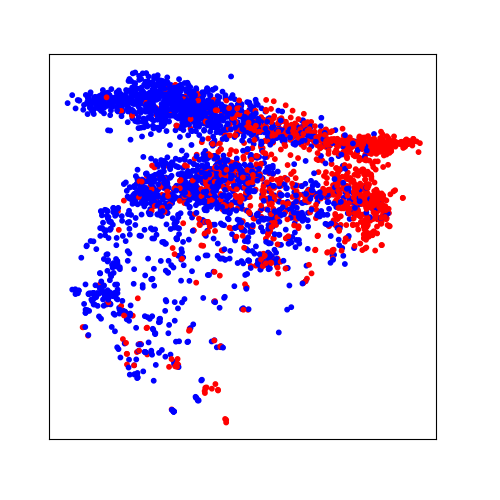}
		\label{MOSEI_student_6}
	}
	\hspace{-0.65cm}
	\subfigure[Student-0.7]{
		\includegraphics[width=0.21\linewidth]{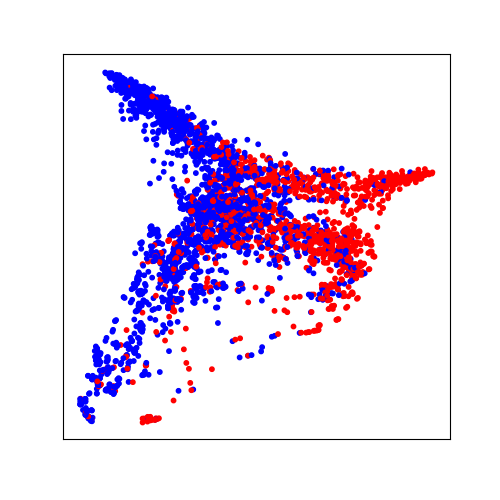}
		\label{MOSEI_student_7}
	}

	\caption{t-SNE visualization results on the CMU-MOSEI test set with the increasing data missing rates. We use red and blue to represent positive and negtive respectively.}
	\label{Embedding Space EI}
\end{figure*}
\begin{figure*}[!h]
	\centering
	\subfigure[Teacher]{
		\includegraphics[width=0.21\linewidth]{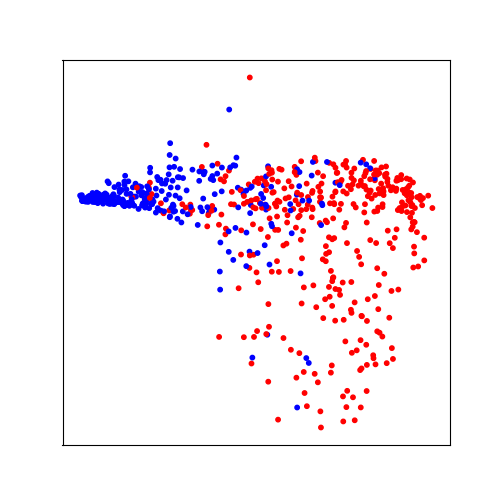}
		\label{MOSI_teacher}
	}
	\hspace{-0.65cm}
	\subfigure[Student-0.0]{
		\includegraphics[width=0.21\linewidth]{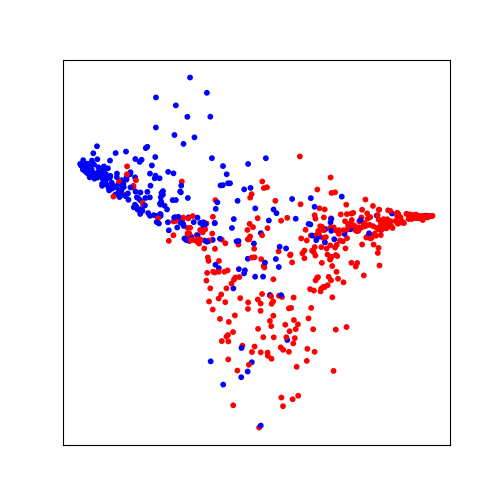}
		\label{MOSI_student_0}
	}
	\hspace{-0.65cm}
	\subfigure[Student-0.1]{
		\includegraphics[width=0.21\linewidth]{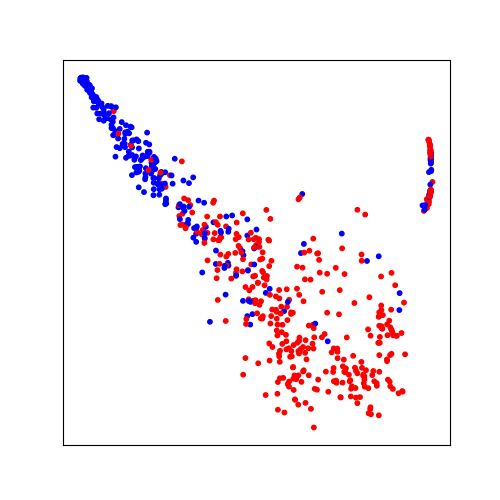}
		\label{MOSI_student_1}
	}
	\hspace{-0.65cm}
	\subfigure[Student-0.2]{
		\includegraphics[width=0.21\linewidth]{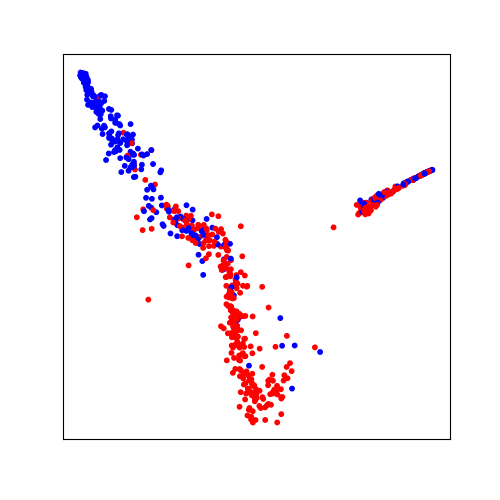}
		\label{MOSI_student_2}
	}
	\hspace{-0.65cm}
	\subfigure[Student-0.3]{
		\includegraphics[width=0.21\linewidth]{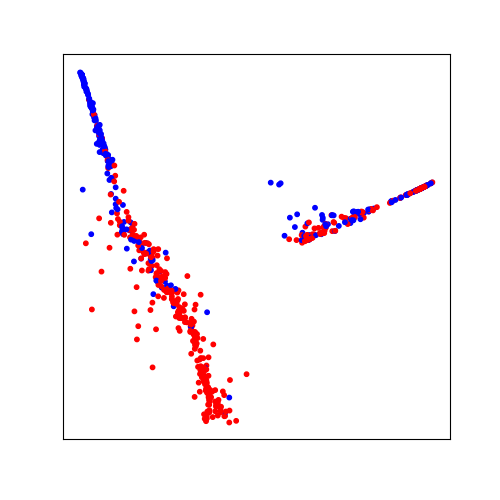}
		\label{MOSI_student_3}
	}
	\hspace{-0.65cm}
	\subfigure[Student-0.4]{
		\includegraphics[width=0.21\linewidth]{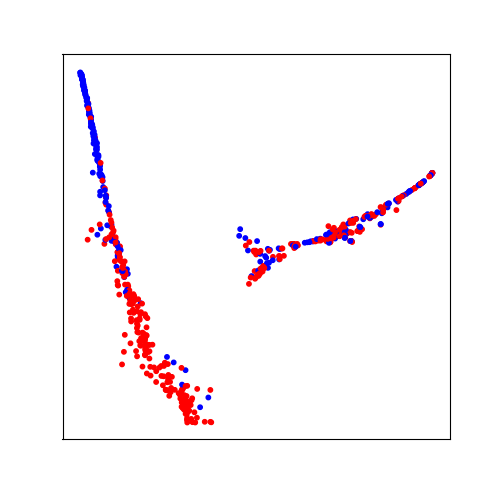}
		\label{MOSI_student_4}
	}
	\hspace{-0.65cm}
	\subfigure[Student-0.5]{
		\includegraphics[width=0.21\linewidth]{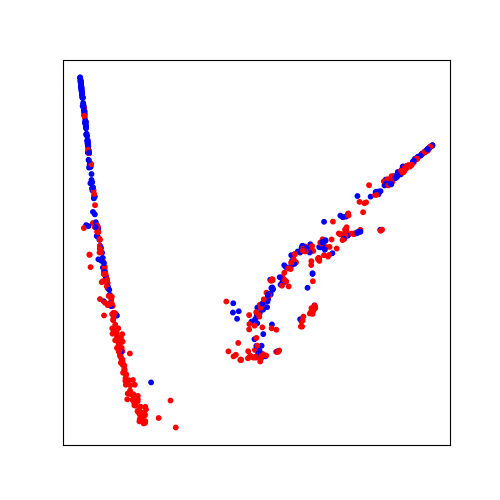}
		\label{MOSI_student_5}
	}
	\hspace{-0.65cm}
	\subfigure[Student-0.6]{
		\includegraphics[width=0.21\linewidth]{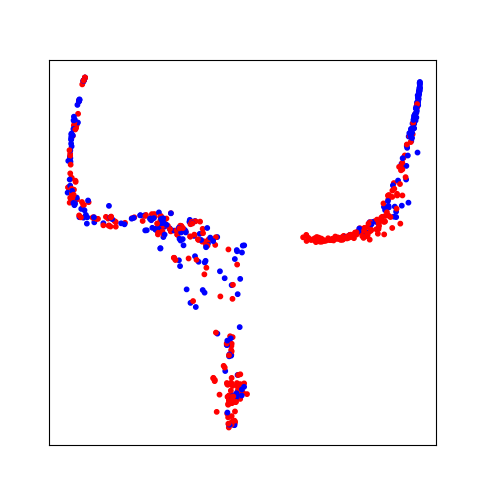}
		\label{MOSI_student_6}
	}
	\hspace{-0.65cm}
	\subfigure[Student-0.7]{
		\includegraphics[width=0.21\linewidth]{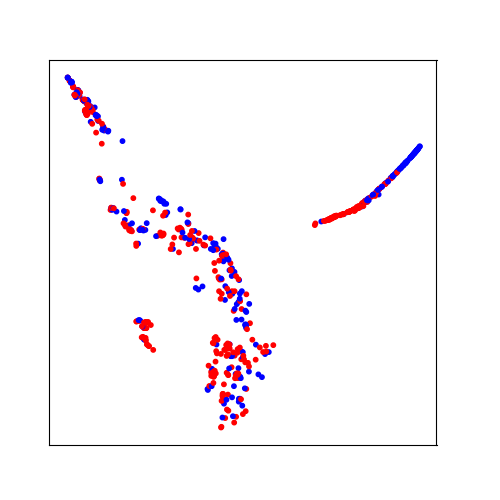}
		\label{MOSI_student_7}
	}

	\caption{t-SNE visualization results on the CMU-MOSI test set with the increasing data missing rates. We use red and blue to represent positive and negtive respectively.}
	\label{Embedding Space SI}
\end{figure*}


\section{Confusion Matrices}
We provide detailed confusion matrix results for the four-class and six-class IEMOCAP tasks. Similar quantities and data distributions at different rates demonstrate the model's stability and robustness when dealing with incomplete data.

\begin{figure*}[!h]
	\centering
	\subfigure[Student-0.0]{
		\includegraphics[width=0.242\linewidth]{imgs/confusion_matrixs/IEMOCAPFour_student_0.0.png}
	}
	\hspace{-0.45cm}
		\subfigure[Student-0.1]{
				\includegraphics[width=0.24\linewidth]{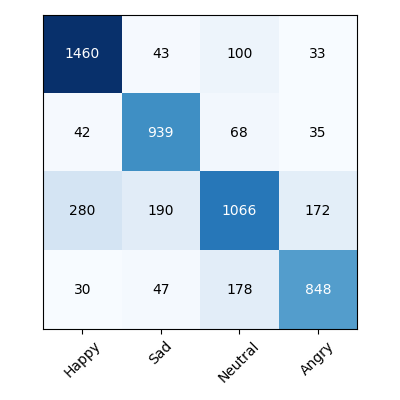}
			}
		\hspace{-0.45cm}
	\subfigure[Student-0.2]{
		\includegraphics[width=0.24\linewidth]{imgs/confusion_matrixs/IEMOCAPFour_student_0.2.png}
	}
	\hspace{-0.45cm}
		\subfigure[Student-0.3]{
				\includegraphics[width=0.24\linewidth]{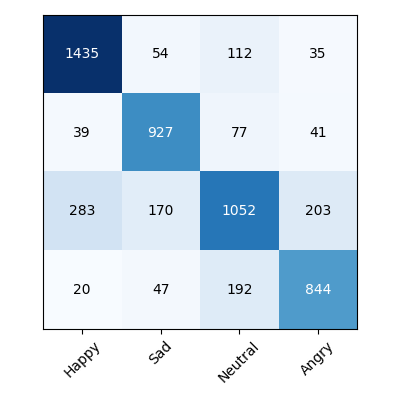}
			}
		\hspace{-0.45cm}
	\subfigure[Student-0.4]{
		\includegraphics[width=0.242\linewidth]{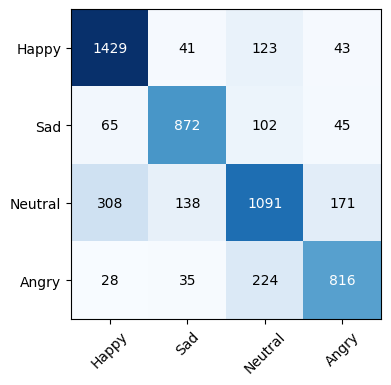}
	}
	\hspace{-0.45cm}
		\subfigure[Student-0.5]{
				\includegraphics[width=0.24\linewidth]{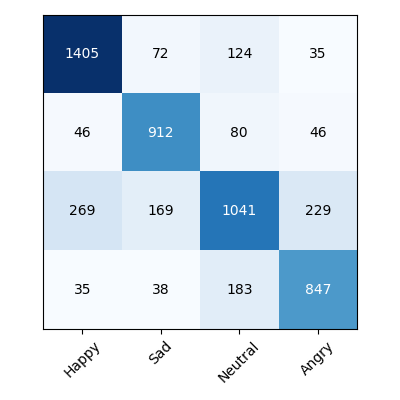}
			}
		\hspace{-0.45cm}
		\subfigure[Student-0.6]{
				\includegraphics[width=0.24\linewidth]{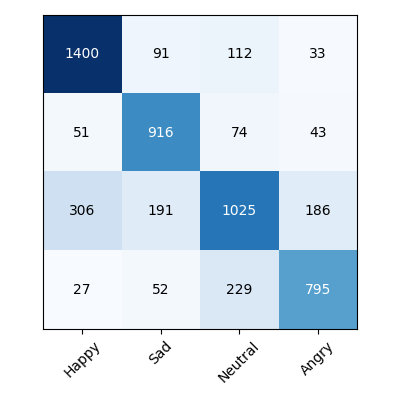}
			}
		\hspace{-0.45cm}
	\subfigure[Student-0.7]{
		\includegraphics[width=0.24\linewidth]{imgs/confusion_matrixs/IEMOCAPFour_student_0.7.png}
		
	}

	\caption{Visualization results of confusion matrices on the four-class IEMOCAP with the increasing data missing rates.}
	\label{FULL Confusion Matrixs}
\end{figure*}

\begin{figure*}[!h]
	\centering
	\subfigure[Student-0.0]{
		\includegraphics[width=0.244\linewidth]{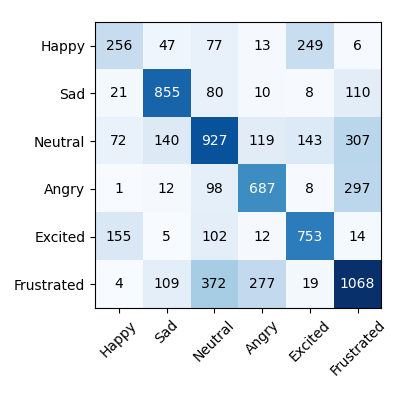}
	}
	\hspace{-0.45cm}
	\subfigure[Student-0.1]{
		\includegraphics[width=0.24\linewidth]{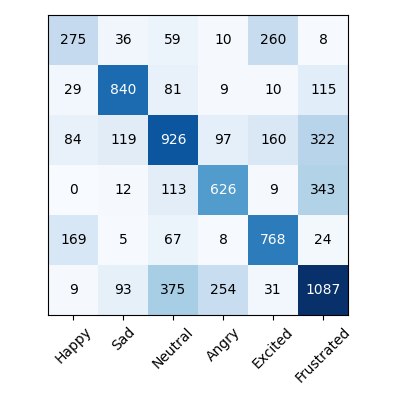}
	}
	\hspace{-0.45cm}
	\subfigure[Student-0.2]{
		\includegraphics[width=0.24\linewidth]{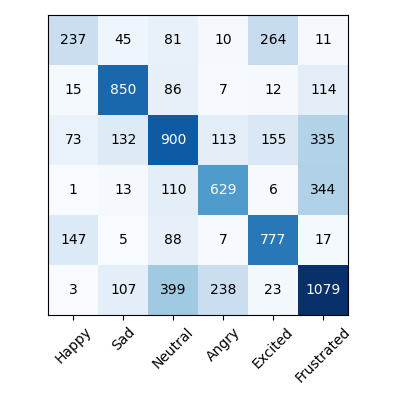}
	}
	\hspace{-0.45cm}
	\subfigure[Student-0.3]{
		\includegraphics[width=0.24\linewidth]{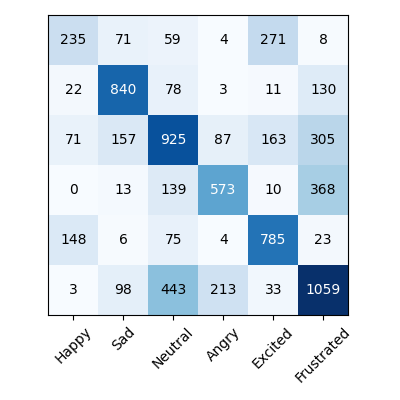}
	}
	\hspace{-0.45cm}
	\subfigure[Student-0.4]{
		\includegraphics[width=0.247\linewidth]{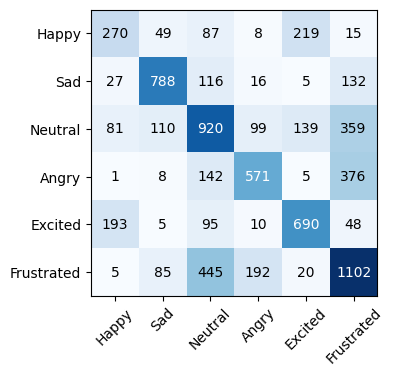}
	}
	\hspace{-0.45cm}
	\subfigure[Student-0.5]{
		\includegraphics[width=0.24\linewidth]{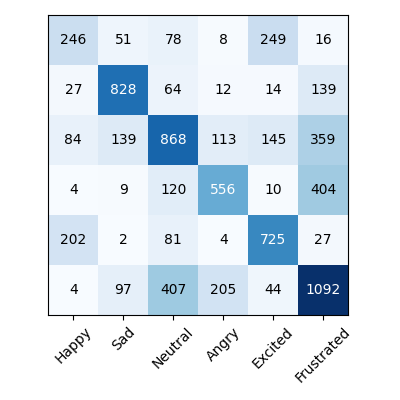}
	}
	\hspace{-0.45cm}
	\subfigure[Student-0.6]{
		\includegraphics[width=0.24\linewidth]{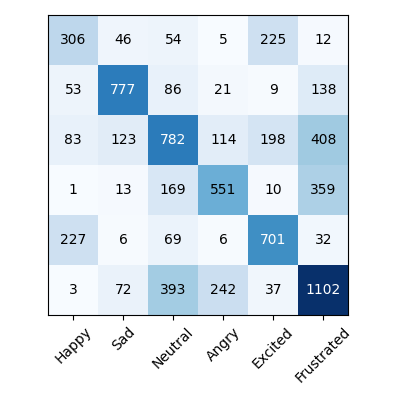}
	}
	\hspace{-0.45cm}
	\subfigure[Student-0.7]{
		\includegraphics[width=0.24\linewidth]{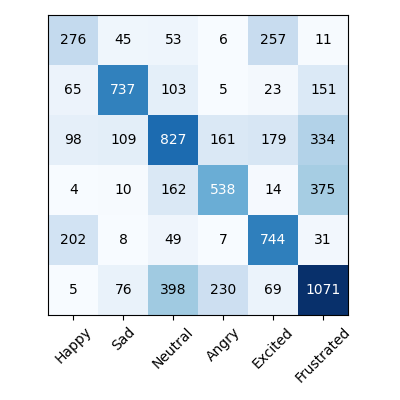}
		
	}

	\caption{Visualization results of confusion matrices on the six-class IEMOCAP with the increasing data missing rates.}
	\label{FULL Confusion Matrixs Six}
\end{figure*}

\section{Search Space}

We select four computationally lightweight operations for our model:
\begin{itemize}
	\item AttentionMixer:
	\begin{equation}
		MixMap = \frac{(X_mW_q)(X_mW_k)^T}{\sqrt{C}}
	\end{equation}
	\begin{equation}
		output = (X_mW_v)\otimes \mathrm{softmax}(MixMap)
	\end{equation}
	$W_q$, $W_k$ and $W_v$ are trainable parameters.
	\item MLPMixer:
	\begin{equation}
		output = MLP(X_m^T)^T
	\end{equation}
	\item AvgPoolMixer:
	\begin{equation}
		output = \mathrm{AvgPool}(X_m^T)^T
	\end{equation}
	\item MaxPoolMixer:
	\begin{equation}
		output = \mathrm{MaxPool}(X_m^T)^T
	\end{equation}
\end{itemize}

\end{document}